\documentclass[%
reprint,
 amsmath,amssymb,
 aps,
 pra,
floatfix,
]{revtex4-2}

\usepackage{graphicx}
\usepackage{dcolumn}
\usepackage{bm}
\usepackage{xcolor}
\usepackage{hyperref}

\begin{document}

\title{Photodetachment energy of negative hydrogen ions}

\author{Maen Salman}
\email{maen.salman@lkb.upmc.fr}
\affiliation{Laboratoire Kastler Brossel, Sorbonne Université, CNRS, ENS-Université PSL,
Collège de France, 4 place Jussieu, F-75005 Paris, France}

\author{Jean-Philippe Karr}
\email{jean-philippe.karr@lkb.upmc.fr}
\affiliation{Laboratoire Kastler Brossel, Sorbonne Université, CNRS, ENS-Université PSL,
Collège de France, 4 place Jussieu, F-75005 Paris, France}
\affiliation{Université Evry Paris-Saclay, Boulevard François Mitterrand,
F-91000 Evry, France}

\date{\today}

\begin{abstract}
We report a high-precision calculation of the photodetachment energy of the hydrogen anion $\mathrm{H}^{-}$, also known as the electron affinity of the hydrogen atom. The nonrelativistic bound-state energy is obtained using an exact three-body approach, and supplemented by leading relativistic, quantum-electrodynamic, finite-nuclear-size, and hyperfine corrections. Our result is $6083.06447(68)\,\mathrm{cm}^{-1}$ for the detachment to the hydrogen ground-state hyperfine level $(F=0)$, which is 220 times more precise than the best experimental determination to date,  $6082.99(15)\,\mathrm{cm}^{-1}$, as reported by Lykke \textit{et al.} Beyond their intrinsic interest, these results provide critical input for antihydrogen physics, where controlled photodetachment of $\bar{\mathrm{H}}^{+}$ offers a path to producing ultracold antihydrogen (and its isotopes) for precision experiments. We further determine the photodetachment thresholds for $^{2}\mathrm{H}^{-}$ and $^{3}\mathrm{H}^{-}$ into the ground hyperfine states of the corresponding hydrogenic atoms, yielding $6086.70679(68)\,\mathrm{cm}^{-1}$ for $^{2}\mathrm{H}(F=1/2)$ and $6087.87924(68)\,\mathrm{cm}^{-1}$ for $^{3}\mathrm{H}(F=0)$.

\end{abstract}

\maketitle
\tableofcontents

\section{Introduction}

\label{sec:Introduction}

The hydrogen anion $\mathrm{H}^{-}$ offers a uniquely clean and stringent test of electron correlation. Owing to its low nuclear charge $Z=1$, it is only weakly bound, and its two electrons exhibit correlation effects that go beyond mean-field theory. In fact, within the nonrelativistic (NR) Hartree--Fock approximation, $\mathrm{H}^{-}$ is predicted to be unstable, with its ground-state energy lying above that of atomic hydrogen, $-0.5\,\mathrm{a.u.}$ This failure is illustrated
by the result of Linderberg (1961) \cite{Linderberg1961}, who derived
the leading coefficients of the $1/Z$ expansion for the ground-state
energy of two-electron atoms within the Hartree--Fock approximation.
For $Z=1$, his results yield $-0.48706\,\mathrm{a.u.}$ A fully converged
numerical Hartree-Fock calculation (all orders in $1/Z$), carried
out by Roothaan and Soukup (1979) \cite{RoothaanSoukup1979}, refined
the value to $-0.487929734372\,\mathrm{a.u.}$, which remains above the hydrogen ground-state energy, thus indicating the absence of a bound state within the Hartree--Fock approximation. This energy value
has since been independently validated by multiple authors \cite{COX2020,Koga1995,Burton2021}.

To improve on these estimates and demonstrate the stability of our system, the decisive factor is explicit inclusion of electron
correlation in the wavefunction. In 1929, Hans Bethe employed the Hylleraas wavefunction \cite{Hylleraas1929}, originally developed for the helium atom and dependent on the inter-electronic distance
$r_{12}$ as an explicit variable, thereby capturing correlations between the two electrons.  Using this \textit{ansatz}, Bethe showed that $\mathrm{H}^{-}$ possesses a stable ground state, estimating its energy to $-0.5253\,\mathrm{a.u.}$ \cite{Bethe1929}. It is worth noting here that Chandrasekhar (1944)
\cite{Chandrasekhar1944} later proposed a very basic trial wavefunction
based on treating the two electrons asymmetrically, a strategy that
indirectly captured the effects of their mutual correlation (\textit{radial
correlation}), without incorporating explicit dependence on the $r_{12}$ variable, and found an energy of $-0.51330\,\mathrm{a.u.}$ However, high accuracy demands
the explicit inclusion of $r_{12}$, which captures the residual angular
correlation between electrons \cite{Schwartz1962,BanyardBaker1969}.

Unlike neutral and positively charged atoms, negative atoms (anions)
possess only a finite number of bound states \cite{Hill1977PhysRevLett.38.643}.
Attempts to identify excited bound states in $\mathrm{H}^{-}$ have not been successful
 \cite{Pekeris1962,Hylleraas1964}, until Hill
demonstrated that, in the infinite proton-mass limit, the system supports
only a single bound state \cite{Hill1977PhysRevLett.38.643}. This
proof was also extended to cover the finite proton mass case, and
Hill proved that the $\mathrm{H}^{-}$ ion has a single bound state
for an electron-to-proton mass ratio of $m/M\lesssim 0.21010636$
\cite{Hill1977_2,Hill1980_3}, which is strongly satisfied for the
actual ratio $m/M\approx5.45\times10^{-4}.$ Equivalently, no bound
state would exist if the proton mass were smaller than $M\lesssim
 4.7594942m$.
Further calculations have established that the critical nuclear charge
required to bind two electrons is $Z\approx0.91103$ \cite{BakerFreundHillNydenMorgan1990,Estienne2014,OlivaresPilonTurbiner2015}.

Many efforts were made on both theoretical and experimental determinations
of the $\mathrm{H}^{-}$ photodetachment threshold, and the goal of
this work is to provide a precise theoretical determination of the
photodetachment energy of $\mathrm{H}^{-}$; the minimum energy required
to remove a single electron. Beyond its fundamental interest,
this quantity is important for the  production of ultracold antihydrogen atoms, for example, in the GBAR experiment \cite{GBAR2015}, which aims to measure the free-fall
acceleration of antihydrogen in Earth's gravity. Indeed, if production of $\bar{\mathrm{H}}^{+}$ ions -- the antimatter counterpart of $\mathrm{H}^{-}$ -- is achieved, these ions can then be trapped and cooled to ultra-low temperatures using sympathetic cooling techniques~\cite{Hilico2014}. Finally, a laser pulse with a well-controlled excess photon energy with respect to the photodetachment threshold, between a few $\mu\mathrm{eV}$ \cite{GBAR2015} and a few tens of $\mu\mathrm{eV}$~\cite{Rousselle2022}, is used to photodetach a single bound positron and produce neutral $\bar{\mathrm{H}}$
atoms with ultra-low kinetic energy. One can estimate that for efficient experimental realization of this scheme, the photodetachment energy needs to be determined with accuracy better than 1 $\mu\mathrm{eV}$, motivating a new independent calculation of this quantity.

In this work, we present a highly accurate estimate of that energy,
based on an accurate nonrelativistic numerical solution of the $\mathrm{H}^{-}$
bound state (including correlation), corrected by recoil, relativistic,
QED, finite nuclear size (FNS), and hyperfine (HF) corrections. The calculations are extended
to the deuterium and tritium negative ions ($^{2}\mathrm{H}^{-}$ and $^{3}\mathrm{H}^{-}$). The present result establishes a new benchmark for photodetachment energies, surpassing
all previous theoretical and experimental determinations in precision.

Over the years, many experiments were conducted to measure the hydrogen anion photodetachment threshold. In Table \ref{tab:tab1_exp}, we list the most significant results. As one can see, the most precise value was provided by Lykke \textit{et al.} in 1991, which agrees with the two subsequent measurements presented in the table. It is worth noting that the experimental accuracy may potentially be improved by a large factor. Photodetachment energies have been measured with about 1 $\mu\mathrm{eV}$ accuracy in several elements using the laser photodetachment microscopy method \cite{Ning2022}, see e.g. \cite{Carette2010}. This is to be compared to the 19 $\mu\mathrm{eV}$ uncertainty of the current best measurement \cite{Lykke1991PhysRevA.43.6104}.

On the theory side, it should be mentioned that highly accurate calculations were reported for light helium-like ions~\cite{Yerokhin2010}, but the $Z=1$ case was not considered. Important theoretical results for the $\mathrm{H}^-$ photodetachment threshold are reported in Table \ref{tab:tab2_theo}. The last three entries of this table, Drake (1988), Kinghorn \& Adamowicz (1997), and Frolov \& Smith (2003), are not independent: the latter two adopt the same total correction of $-0.307\,505\,\mathrm{cm}^{-1}$
from Drake, comprising relativistic, relativistic-recoil, Lamb shift,
and finite nuclear size contributions. None of these works reports error bars. This absence of independent verification and quantified
uncertainties is a primary motivation for the present study.

The organization of the paper is as follows. Sec. \ref{sec:Theory}
presents the theoretical framework for hydrogenic (hydrogen isotopes)
and helium-like systems (negative hydrogen isotopes). Numerical evaluations
for neutral hydrogen isotopes and their corresponding negative ions
are reported in Sec. \ref{sec:Numerical-evaluations}. The resulting
photodetachment energies are discussed in Sec. \ref{subsec:Photodetachment-energy-of}
and compared with previous theoretical predictions and experimental
determinations. A summary and outlook are given in Sec. \ref{sec:Summary-and-outlook}.
Details of the high-precision evaluation of the Bethe logarithm are
provided in Appendix \ref{sec:Many-body-Bethe-logarithm}.

Throughout the manuscript, equations are expressed in atomic units
($e=\hbar=4\pi\epsilon_{0}=1$, $\alpha\equiv e^{2}/4\pi\epsilon_{0}\hbar c=c^{-1}$),
except for the electron mass $m$, which is retained to provide clarity
in expressions involving electron and nuclear masses.

\begin{table}[htbp]
\centering
\begin{tabular}{|l|c|l|l|}
\hline 
Author & Year & Value in $\mathrm{eV}$ & Ref.\tabularnewline
\hline 
\hline 
Khvostenko and Dukel'skii & 1960 & $0.8(1)$ & \cite{khvostenko1960formation}\tabularnewline
\hline 
Armstrong & 1963 & $0.745$ & \cite{Armstrong1963PhysRev.131.1132}\tabularnewline
\hline 
Weisner and Armstrong & 1964 & $0.77(2)$ & \cite{Weisner_Armstrong_1964}\tabularnewline
\hline 
Branscomb & 1967 & $0.7563(62)$ & \cite{borowitz1968physics}\tabularnewline
\hline 
Berry & 1969 & $0.756(13)$ & \cite{Berry1969}\tabularnewline
\hline 
Feldmann & 1970 & $0.776(20)$ & \cite{Feldmann1970}\tabularnewline
\hline 
McCulloh and Walker & 1974 & $\geq0.754(2)$ & \cite{MCCULLOH1974}\tabularnewline
\hline 
Feldmann & 1975 & $0.7539(20)$ & \cite{FELDMANN1975}\tabularnewline
\hline 
Popp and Kruse & 1976 & $0.775(24)$ & \cite{POPP1976,POPP1975}\tabularnewline
\hline 
Scherk & 1979 & $0.7451(41)$ & \cite{Scherk1979}\tabularnewline
\hline 
Donahue \textit{et al.} with Frost & 1980 & $0.753(5)$ & \cite{DonahueGramHammHammBryantButterfieldClarkFrostSmith1981,Frost1981}\tabularnewline
\hline 
Lykke \textit{et al.} & 1991 & $0.754\,195(19)$ & \cite{Lykke1991PhysRevA.43.6104}\tabularnewline
\hline 
Harms \textit{et al.} & 1997 & $0.754\,171(87)$ & \cite{Oliver_Harms_1997}\tabularnewline
\hline 
Beyer and Merkt & 2018 & $0.754\,270(62)$ & \cite{BeyerMerkt2018}\tabularnewline
\hline 
\end{tabular}
\caption{\label{tab:tab1_exp}Selection of experimental values of the $\mathrm{H}^{-}$ photodetachment threshold.}
\end{table}

\begin{table}[htbp]
\centering
\begin{tabular}{|c|l|l|l|}
\hline 
Year & Value in $\mathrm{cm}^{-1}$ & Corrections & Ref.\tabularnewline
\hline 
\hline 
1957 & $6060.03$ & MP1 & \cite{HartHerzberg1957PhysRev.106.79}\tabularnewline
\hline 
1958 & $6083.08$ & MP1, Rel, Lamb & \cite{Pekeris1958}\tabularnewline
\hline 
1962 & $6083.0958$ & MP1, Rel & \cite{Pekeris1962}\tabularnewline
\hline 
1970 & $6083.13$ & MP1, Rel, Lamb & \cite{AASHAMAR1970}\tabularnewline
\hline 
1988 & $6083.099414$ & MPa, Rel+rec, Lamb, FNS & \cite{DRAKE1988}\tabularnewline
\hline 
1997 & $6083.0994$ & MPa, Rel+rec, Lamb, FNS & \cite{KinghornAdamowicz1997}\tabularnewline
\hline 
2003 & $6083.09937$ & MPa, Rel+rec, Lamb, FNS & \cite{FrolovSmith2003}\tabularnewline
\hline 
\end{tabular}
\caption{\label{tab:tab2_theo}Main theoretical predictions of the
$\mathrm{H}^{-}$ photodetachment threshold. MP1 and MPa denote the
first-order and all-order mass polarization corrections, respectively.
Rel is the leading-order relativistic correction; Rel+rec includes
its recoil correction. Lamb represents the leading QED correction,
including the Bethe logarithm and vacuum polarization. FNS is the
finite nuclear size correction.}
\end{table}

\section{Theory}
\label{sec:Theory}

We present the theoretical framework for both hydrogen- and helium-like systems, along with their leading-order corrections, as the photodetachment energy is defined by the ground-state energy difference between them. Corresponding numerical evaluations of the NR ground states, and their corrections, are reported in Sec. \ref{sec:Numerical-evaluations}.

\subsection{Neutral hydrogen isotopes}
\label{subsec:Hydrogenic-section}

For the hydrogen problem, we start with the two-body Schr{\"o}dinger equation, which, in the center-of-mass frame, reduces to
\begin{equation} H^{(0)}\psi(\mathbf{r})=E^{(0)}\psi(\mathbf{r}),\quad\text{with}\quad H^{(0)}=\frac{\hat{\mathbf{p}}^2}{2\mu}-\frac{Z}{r},\label{eq:H1} \end{equation}
where $\mu=mM/(m+M)$ is the reduced electron mass, with $m$ and $M$
the electron and nuclear mass, respectively. The energy levels
are classified by the principal quantum number $n$,
\begin{equation}
E^{(0)}=-\frac{1}{2}\mu(Z/n)^{2}.\label{eq:ENR}
\end{equation}
We adopt the notation $E^{(n)}$ to denote contributions of order
$\alpha^{n+2}mc^{2}$, or equivalently $\alpha^{n}m$
in our system of units. The leading-order relativistic correction
incorporating recoil effects was first evaluated by Barker and Glover
in 1955 \cite{BarkerGlover1955PhysRev.99.317}. If one neglects terms
associated with anomalous magnetic moments and the hyperfine correction
(to be considered later), one finds the following energy shift
\begin{equation}
\begin{aligned}E^{(2)}& =-\frac{1}{2}\alpha^{2}\mu\frac{Z^{4}}{n^{3}}\bigg[\frac{1}{j+1/2}-\frac{3}{4n}+\frac{\mu}{4n(m+M)}\bigg]\\
 & +\frac{1}{2}\alpha^{2}\mu\frac{Z^{4}}{n^{3}}\big(\frac{\mu}{M}\big)^{2}\!\bigg[\frac{1}{j+1/2}\!-\!\frac{1}{\ell+1/2}\bigg](1\!-\!\delta_{\ell,0}),
\end{aligned}
\label{eq:EREL}
\end{equation}
a form which first appeared in the work of Sapirstein and Yennie \cite{SapirsteinYennie1990}.
For further discussions about this correction, the reader may consult
Ref. \cite{Grotch1967}, \cite[section 12.5]{JentschuraAdkins2022book},
and \cite[section 3.1]{Eides2007}. It should be noted that the Kronecker
delta term $\delta_{\ell,0}$ must be omitted when the nucleus is
a spin-$0$ particle (charged scalar boson, e.g., ${}^{4}\mathrm{He}$ nucleus)
\cite[section 9.7]{BjorkenDrell1964}\cite[section 5.1]{Owen1994},
or a spin-$1$ particle (charged vector boson, e.g., the deuteron) \cite{PachuckiKarshenboim1995}.

The most important QED correction comes from the leading-order self-energy
correction, and reads \cite[Eqs. (4.352, 4.353)]{JentschuraAdkins2022book}
\begin{equation}
\begin{aligned}E^{\mathrm{SE}} & =\frac{4}{3}\alpha^{3}m\frac{Z^{4}}{\pi n^{3}}\left(\frac{\mu}{m}\right)^{3} \bigg[ \big\{\ln\big(\frac{m}{\mu(\alpha Z)^{2}}\big)+\frac{11}{24}+\frac{3}{8}\\
 & \,\,\,\quad -\ln(k_{0}/(Z^{2}\mathrm{Ry}_{\!M}))\big\} \delta_{\ell,0}-\frac{3}{8}\frac{1-\delta_{\ell,0}}{\kappa(2\ell+1)}\frac{m}{\mu } \bigg] ,
\end{aligned}\label{eq:ESE}
\end{equation}
where $\mathrm{Ry}_{\!M}=(\mu/m)\mathrm{Ry}$ is the reduced-mass Rydberg energy and $\mathrm{Ry}=m/2$ is the Rydberg energy unit. The $+3/8$ factor in the first line represents the effect of the electron anomalous magnetic moment. The quantity  $\kappa=(-1)^{j+\ell+\frac{1}{2}}(j+1/2)$ is the relativistic angular quantum number.  The Bethe logarithm $\delta_{\ell,0}\ln(k_{0}/(Z^{2}\mathrm{Ry}_{\!M}))$ associated with an eigensolution $(E^{(0)}_{0},\psi_{0})$ of Eq.~(\ref{eq:H1}) is given by the following sum over all eigensolution $(E^{(0)}_{i},\psi_{i})$ of the same Hamiltonian, as \cite[Eqs. (4.318)]{JentschuraAdkins2022book}
\begin{equation}
\begin{aligned} &\delta_{\ell,0}\ln(k_{0}/(Z^{2}\mathrm{Ry}_{\!M}))\\
& =\frac{n^{3}}{2Z^{4}}\sum_{i}|\langle i |\hat{\mathbf{p}}/\mu|0\rangle|^{2}\frac{E^{(0)}_{i}\!-E^{(0)}_{0}}{\mu}\!\ln\!\big(\!\frac{|E^{(0)}_{i}\!-E^{(0)}_{0}|}{Z^{2}\mathrm{Ry}_{\!M}}\!\big),
\end{aligned}
\label{eq:k0_hydrogen}
\end{equation}
where $\psi_i(\mathbf{r}) = \langle \mathbf{r}|i\rangle$ is the position-space representation of the state $|i\rangle$, and $n$ is the principal quantum number of the $| 0\rangle$ state. With the factor $Z^{2}(\mu/m)\mathrm{Ry}$ in the logarithm denominator, the Bethe logarithm is independent of the reduced mass $\mu$ and the nuclear charge $Z$. The second most important QED correction is
the leading-order vacuum polarization contribution, which only affects
$s$ states, and can be shown to read \cite[Eq. (11.164)]{JentschuraAdkins2022book}
\begin{align}
E^{\mathrm{VP}} & =-\frac{4}{15}\alpha^{3}m\frac{Z^{4}}{\pi n^{3}}\left(\frac{\mu}{m}\right)^{3}\delta_{\ell,0},\label{eq:EVP}
\end{align}
as first derived by Dirac \cite{dirac1934} and Heisenberg \cite{Heisenberg1934}.
The $(\mu/m)^{3}$ factor appearing here, as in the self-energy and
the next two corrections, originates from the dependence of these
terms on the value of the electron's probability density at the origin. At the same order in $\alpha$, comes the relativistic recoil correction
(Salpeter correction), first calculated for $n=2$ by Salpeter \cite{Salpeter1952PhysRev.87.328},
and re-checked by Fulton and Martin \cite{FultonMartin1954PhysRev.95.811}.
The general $n$ problem was evaluated by Erickson and Yennie \cite{EricksonYennie1965},
and later verified by Bhatt and Grotch \cite{BhattGrotch1985}. For
modern detailed derivations of this correction one may consult Ref.
\cite[section 15.5]{JentschuraAdkins2022book} and \cite[section 4.1]{Eides2007}.
Results of the last six references concerned the interaction between
two spin-$1/2$ particles, with different masses and charges. These
results were extended to include the case of a spin-$1/2$ particle interacting
with a spin-$0$ or spin-$1$ particles \cite{Shelyuto2018,Shelyuto2019},
to obtain the following general expression \cite[Eq. (54)]{Pachucki2024}
\begin{equation}
\begin{aligned}& E^{(3),\mathrm{RC}} =\\
 & \quad \alpha^{3}m\frac{Z^{5}}{\pi n^{3}}\left(\frac{\mu}{m}\right)^{3}\frac{m}{M}\bigg(\bigg\{\frac{2}{3}\ln(Z\alpha)^{-1}\\
 & +\!\frac{14}{3}\big[\ln\big(\frac{2}{n}\big)+\psi(n+1)-\psi(1)+\frac{2n-1}{2n}\big]-\frac{1}{9}\\
 & -\!2\ln\big(1\!+\!\frac{m}{M}\big)+\frac{m^{2}}{M^{2}\!-\!m^{2}}\ln\left(\!\frac{M}{m}\!\right)\big[2\!+\!I(2I\!-\!1)\big]\\
 & -\!\frac{8}{3}\!\ln(k_{0}/(Z^{2}\mathrm{Ry}_{\!M}))\!\bigg\}\delta_{\ell,0}-\frac{7}{3}\frac{1-\delta_{\ell,0}}{\ell(\ell+1)(2\ell+1)}\bigg),
\end{aligned}
\end{equation}
where $\psi(n)$ denotes the digamma function, and $I$ is the nuclear spin: $I=1/2$ for the proton and triton, and $I=1$ for the deuteron. The next-order term in the recoil expansion
is given by \cite[Eq. (5.8)]{FultonMartin1954PhysRev.95.811}\cite[Eq. (2.6a)]{SapirsteinYennie1990}
\begin{equation}
\begin{aligned} & E^{(3),\mathrm{NSE}}=\frac{4}{3}\alpha^{3}m\frac{Z^{6}}{\pi n^{3}}\left(\frac{\mu}{m}\right)^{3}\left(\frac{m}{M}\right)^{2}\\
 & \quad\quad\times\big[\ln\big(\frac{M}{\mu(\alpha Z)^{2}}\big)\!+\!\frac{5}{6}\!-\!\ln(k_{0}/(Z^{2}\mathrm{Ry}_{\!M}))\big]\delta_{\ell,0},
\end{aligned}\label{eq:NSE}
\end{equation}
representing the nuclear self-energy process. The last four corrections form the total $\alpha^{3}$ energy shift 
\begin{equation}
E_{\mathrm{total}}^{(3)}=E^{\mathrm{SE}}+E^{\mathrm{VP}}+E^{(3),\mathrm{RC}}+E^{(3),\mathrm{NSE}}.
\end{equation}

We draw attention to an important point concerning $E_{n,\ell}^{(3),\mathrm{NSE}}$.
As noted by Pachucki, there exists an inherent ambiguity in the separation
between the nuclear charge radius contribution, see Eq. (\ref{eq:EFNS}) below,
and the nuclear self-energy term in Eq. (\ref{eq:NSE}). This ambiguity
is associated with the freedom to absorb constant terms into the definition
of the nuclear radius. In the present work, it manifests itself through
the constant $+5/6$ appearing in Eq. (\ref{eq:NSE}) \cite{Pachucki1995}\cite[section 5.6]{YerokhinPachuckiPatkos2019}. At the level of precision
sought in the present work, this arbitrariness has no numerical impact,
as well as the whole $E_{n,\ell}^{(3),\mathrm{NSE}}$ term. Additional
discussions of this issue can be found in Refs. \cite{Jentschura2011},
\cite[section 5.1.3]{Eides2007}, and \cite[section III.L.]{Pachucki2024}.

The next contribution we include arises at order $\alpha^{4}m$, corresponding
to second-order relativistic and QED corrections. It can be expressed
as
\begin{equation}
E^{(4)}=E_{\mathrm{Rel}}^{(4)}+E_{R_{1}}^{(4)}+E_{R_{2}}^{(4)},
\end{equation}
where $E_{\mathrm{Rel}}^{(4)}$ denotes the $\alpha^{4}m$ relativistic
term, obtained from the $Z\alpha$-expansion of the hydrogenic Dirac energy. The term $E_{R_{1}}^{(4)}$ represents
the $\alpha^{4}m$ one-loop radiative correction, corresponding to
the $A_{50}$ coefficient in the $Z\alpha$-expansion of the self-energy
and vacuum polarization. Finally, $E_{R_{2}}^{(4)}$ accounts for
the leading two-loop QED contribution, associated with the $B_{40}$
coefficient, which collects the SESE, VPVP, and SEVP diagrams (see Ref.~\cite{YerokhinPachuckiPatkos2019}). These
terms are explicitly given by
\begin{align}
E_{\mathrm{Rel}}^{(4)} & =-\alpha^{4}m\frac{Z^{6}}{16}+{\cal{O}}(\frac{m}{M})^2,\,\, \text{for}\,\, (n,\kappa)=(1,-1),\label{eq:E4rel}\\
E_{R_{1}}^{(4)} & =\alpha^{4}m\frac{Z^{5}}{n^{3}} \left(\frac{\mu}{m}\right)^{3}  \bigg[\frac{427}{96}-2\ln(2)\bigg] \delta_{\ell0}\label{eq:E4r1}\\
E_{R_{2}}^{(4)} & =\alpha^{4}m\frac{Z^{4}}{n^{3}} \left(\frac{\mu}{m}\right)^{3}  \bigg[\frac{3\ln(2)}{2} -\frac{9\zeta(3)}{4\pi^{2}} \nonumber\\ 
& - \frac{2179}{648\pi^{2}} - \frac{10}{27}\bigg] \delta_{\ell0},\label{eq:E4r2}
\end{align}
which can be collected from Refs.~\cite[section 3.1]{Eides2007}~\cite[Eqs. (7 and 16)]{YerokhinPachuckiPatkos2019}~\cite[Eqs. (34 and 35)]{Yelkhovsky2001}.

Although numerically small, the finite-nuclear-size (FNS) correction is included for completeness. It is given by
\cite[\textsection 120]{LandauLifshitz1977}
\begin{align}
E^{\mathrm{FNS}} & =\frac{2}{3}\alpha^{2}mZ^{4}(r_{\mathrm{N}}/\lambdabar)^{2}\frac{1}{n^{3}}\left(\frac{\mu}{m}\right)^{3}\delta_{\ell,0},\label{eq:EFNS}
\end{align}
and similarly to the VP correction, only affects $s$ states. $\lambdabar=\hbar/mc$
is the reduced Compton wavelength and $r_{\mathrm{N}}$ is the root-mean-square
charge radius of the nucleus, extracted from experimental data, following
the definition of the Sachs electric form factor.

The final correction required at our level of precision arises from the hyperfine interaction, first introduced by Fermi~\cite{Fermi1930}.
This interaction yields the following energy
shift \cite{BetheSalpeter1957}
\begin{align}
E^{\mathrm{HF}} & ={\cal A}\frac{3}{8n^{3}}\frac{F(F+1)-I(I+1)-j(j+1)}{j(j+1)(2\ell+1)},\label{eq:EHFS}
\end{align}
where $F$ is the total angular momentum quantum number, with $\mathbf{F} = \mathbf{I}+\mathbf{j}$. Taking only the two leading orders in $\alpha$ into account, the energy prefactor is ${\cal A}=\alpha^{2}mZ^{3}(2/3)g_{\mathrm{e}}g_{\mathrm{N}}(m/M_{p})(\mu/m)^{3}=E_{F}(1+a_{e})$, $g_{\mathrm{e}}$ and $g_{\mathrm{N}}$ are the electron and nucleus g-factors,
respectively, $E_{F}$ is the Fermi energy, and $a_{\mathrm{e}}=g_{\mathrm{e}}/2-1$
is the anomalous magnetic moment. Accounting
for the finite nuclear mass correction yields the $(\mu/m)^{3}$ factor
present in ${\cal A}$, as first noted by Breit and Meyerott \cite{BreitMeyerott1947}. 

Having established the hydrogenic case, we now turn to the helium-like problem and present the corresponding contributions.

\subsection{Negative hydrogen isotopes}

\label{subsec:Helium-like-section}

We shall consider the helium-like system, which is essentially a three-body
system. We shall use $\mathbf{R}_{1}$, $\mathbf{R}_{2}$
and $\mathbf{R}_{0}$, and $\hat{\mathbf{P}}_{1}$, $\hat{\mathbf{P}}_{2}$
and $\hat{\mathbf{P}}_{0}$ to represent position and momentum
operators for the two electrons, and the single nucleus, respectively.
The three-body equation of the helium-like problem reads
\begin{equation}
\begin{aligned} & H^{(0)}\psi(\mathbf{R}_{0},\mathbf{R}_{1},\mathbf{R}_{2})=E^{(0)}\psi(\mathbf{R}_{0},\mathbf{R}_{1},\mathbf{R}_{2})\\
 & H^{(0)}=\frac{\hat{\mathbf{P}}_{0}^{2}}{2M}\!+\!\sum_{i=1}^{2}\!\frac{\hat{\mathbf{P}}_{i}^{2}}{2m}\!-\!\sum_{i=1}^{2}\!\frac{Z}{|\mathbf{R}_{i}\!-\!\mathbf{R}_{0}|}\!+\!\frac{1}{|\mathbf{R}_{1}\!-\!\mathbf{R}_{2}|}.
\end{aligned}
\label{eq:3body}
\end{equation}

In the center-of-mass frame, the corresponding Schr{\"o}dinger equation
can be shown to reduce to \cite[section 11.1]{Drake2006}
\begin{equation}
\begin{aligned} & \tilde{H}^{(0)}\tilde{\psi}(\mathbf{r}_{1},\mathbf{r}_{2})=\tilde{E}^{(0)}\tilde{\psi}(\mathbf{r}_{1},\mathbf{r}_{2}),\\
 & \tilde{H}^{(0)}=\sum_{i=1}^{2}\frac{\hat{\mathbf{p}}_{i}^{2}}{2\mu}+\frac{1}{M}\hat{\mathbf{p}}_{1}\cdot\hat{\mathbf{p}}_{2}-\sum_{i=1}^{2}\frac{Z}{r_{i}}+\frac{1}{r_{12}},\label{eq:3body2body}
\end{aligned}
\end{equation}
where we employ relative coordinates, given by $\mathbf{r}_{i}=\mathbf{R}_{i}-\mathbf{R}_{0}$,
for $i=1,2$, and $\mathbf{r}_{12}=\mathbf{r}_{1}-\mathbf{r}_{2}=\mathbf{R}_{1}-\mathbf{R}_{2}$.
Differential operators $\hat{\mathbf{p}}_{i}$ are associated
with the relative lowercase coordinates. As one can see, in the infinite
nuclear mass $M$ limit, the third term (mass-polarization) vanishes,
and $\mu\rightarrow m$. In Sec. \ref{sec:Numerical-evaluations}
below, we shall numerically solve this last equation for both finite
and infinite nuclear mass $M$.

In 1961, Schwartz considered the helium atom problem, where a state
associated with $(L,M_{L})$ can be written as
\begin{equation}
\tilde{\psi}_{L,M_{L}}(\mathbf{r}_{1},\mathbf{r}_{2})=\sum_{\ell_{1},\ell_{2}=0}^{\infty}F_{\ell_{1},\ell_{2}}^{L}(r_{1},r_{2}){\cal Y}_{L,M_{L}}^{\ell_{1},\ell_{2}}(\hat{\mathbf{r}}_{1},\hat{\mathbf{r}}_{2}),
\end{equation}
where ${\cal Y}_{L,M_{L}}^{\ell_{1},\ell_{2}}$ is the spherical biharmonic
function, discussed in Refs. \cite[section 5.16.1]{Varshalovich1988}
and \cite[chapter X]{CohenTannoudjiDiuLaloe2020}. An important
complexity arises from the fact that this equation includes infinite
sums over $\ell_{1}$ and $\ell_{2}$, and Schwartz looked for the
following alternative expression of this last equation \cite[Eq. (A.4)]{Schwartz1961}
\begin{equation}
\tilde{\psi}_{L,M_{L}}\!(\mathbf{r}_{1},\!\mathbf{r}_{2})=\!\!\!\!\sum_{\{\ell_{1},\ell_{2}\}}\!\!{\cal F}_{\ell_{1},\ell_{2}}^{L}(r_{1},r_{2},r_{12}){\cal Y}_{L,M_{L}}^{\ell_{1},\ell_{2}}\!(\hat{\mathbf{r}}_{1},\!\hat{\mathbf{r}}_{2}),
\end{equation}
where the $r_{12}$ distance is absorbed by the new radial function ${\cal F}_{\ell_{1},\ell_{2}}^{L}$,
and the infinite summation is reduced to a sum over a finite set of
individual angular momenta $\ell_{1}$ and $\ell_{2}$, $\{\ell_{1},\ell_{2}\}$.
Schwartz showed that for a particular $L$, this set is restricted
by the relation $\ell_{1}+\ell_{2}=L+\varpi,$ where $\varpi$ depends on parity ($\varpi=0$ for natural, and $1$ for unnatural). Finally, in order
to guarantee that the overall wavefunction is antisymmetric under the
exchange of particles $1$ and $2$, we write the final expression as
\begin{equation}
\begin{aligned} & \tilde{\psi}_{L,M_{L}}^{\varpi,S,M_{S}}(\mathbf{r}_{1},\mathbf{r}_{2})\\
 &\!=\left(1\!+\!(-\!1)^{S}{\cal P}\right)\!\sum_{\ell_{1}=\varpi}^{L}\!{\cal F}_{\ell_{1},\ell_{2}}^{L}(r_{1},r_{2},r_{12}){\cal Y}_{L,M_{L}}^{\ell_{1},\ell_{2}}{\cal S}_{S,M_{S}},
\end{aligned}
\label{eq:3bodywv}
\end{equation}
with $\ell_2=L+\varpi-\ell_1$. In this expression, ${\cal P}$ is the exchange operator, swapping $\mathbf{r}_{1}\leftrightarrow\mathbf{r}_{2}$,
and $S$ is the total spin, guaranteeing for a singlet (triplet) state
a symmetric (antisymmetric) spatial wavefunction. The total spin state
${\cal S}_{S,M_{S}}$ can be written as
\begin{equation}
{\cal S}_{S,M_{S}}\!=\!\sum_{m_{1},m_{2}}\chi_{m_{1}}\!\otimes\!\chi_{m_{2}}\langle1/2,1/2,m_{1},m_{2}|S,M_{S}\rangle,
\end{equation}
where $\chi_{+1/2}\!=\!\left[\begin{smallmatrix}1\\ 0 \end{smallmatrix}\right]$ and $\chi_{-1/2}\!=\!\left[\begin{smallmatrix}0\\ 1 \end{smallmatrix}\right]$ are spin-$1/2$ eigenstates. 

The goal is then to compute the complicated and unknown radial function ${\cal F}_{\ell_{1},\ell_{2}}^{L}(r_{1},r_{2},r_{12})$.
These calculations are carried out variationally, and the most effective
schemes employ basis functions that depend explicitly on the interparticle
distance $r_{12}$, ensuring a compact and rapidly convergent description
of electron correlation. Two major prescriptions are commonly used.
The first one consists of expanding the radial function in powers
of $r_{1}$, $r_{2}$, and $r_{12}$, as \cite{Drake2006}
\begin{equation}
\begin{aligned} & {\cal F}_{\ell_{1},\ell_{2}}^{L}(r_{1},r_{2},r_{12})\\
 & \approx r_{1}^{\ell_{1}}r_{2}^{\ell_{2}}\sum_{i=0}^{m}\sum_{j=0}^{m-i}\sum_{k=0}^{m-i-j}a_{ijk}r_{1}^{i}r_{2}^{j}r_{12}^{k}e^{-\alpha r_{1}-\beta r_{2}},
\end{aligned}\label{eq:Hylleraas}
\end{equation}
where the (typically real) exponential parameters $\alpha$ and $\beta$
can be optimized to minimize the energy of the sought state, and $a_{ijk}$
are the linear coefficients to be determined by the diagonalization
procedure of the corresponding generalized eigenvalue problem. This basis is usually referred to as the Hylleraas basis. The original Hylleraas basis was introduced using trial wave functions in the form of expansions in the coordinates $s=r_{1}+r_{2}$, $t=r_{2}-r_{1}$ and $u=r_{12}$, together with a similar exponential factor \cite{Hylleraas1929,Hylleraas1930}.
The pre-selected positive integer $m$ (last expression) truncates
the radial expansions and includes all power terms satisfying $i+j+k\leq m$
(\textit{Pekeris shell}), and with a large enough $m$, convergence
is achieved. For extra information about the basis construction, we
refer to the works of Drake and coworkers \cite{Drake1994,DrakeCassarNistor2002,DRAKE2004,PetrimoulxBondyEneSatiDrake2025}.
These works explored double and triple basis sets, where additional
expansions are appended to Eq. (\ref{eq:Hylleraas}) with new sets
$\{\alpha_{2},\beta_{2}\}$ and $\{\alpha_{3},\beta_{3}\}$, optimized
to accelerate energy convergence. 

In the second approach, one expands the radial function as \cite{ThakkarSmith1977,FrolovSmith1995}
\begin{align}
{\cal F}_{\ell_{1},\ell_{2}}^{L}(r_{1},r_{2},r_{12}) & \approx r_{1}^{\ell_{1}}r_{2}^{\ell_{2}}\sum_{i=1}^{N}a_{i}e^{-\alpha_{i}r_{1}-\beta_{i}r_{2}-\gamma_{i}r_{12}},\label{eq:Thakkar}
\end{align}
where $\alpha_{i}$, $\beta_{i}$, and $\gamma_{i}$ may be real or, more generally, complex. This basis was extensively used
by both Korobov \cite{Korobov2000,Korobov2025} and Frolov \cite{FrolovSmith1995,Frolov2004}.
The absence of power expansion in the three radial variables (c.f.
Eq. (\ref{eq:Hylleraas})) is compensated by introducing a large set of
($3N$) exponents, typically generated through a pseudo-random procedure,
presented below.

Both approaches capture leading radial powers and the exponential
decaying behavior in the $r_{1}$ and $r_{2}$ parameters. Each basis
construction has its advantages and limitations. Although the second
approach is simpler, it requires high numerical precision to avoid
linear dependencies for large $N$. However, it allows flexibility
in optimizing the $\alpha$, $\beta$, and $\gamma$ exponents to
converge to a lower upper bound for the nonrelativistic energy. Historically,
various basis constructions have been proposed, including integer
and fractional powers of $r_{1}$, $r_{2}$, and $r_{12}$, or the
Hylleraas variables $s$, $t$, and $u$, with or without logarithmic
functions of these variables. In Ref. \cite{Schwartz2006Review},
Schwartz provided a comprehensive comparison of different basis set
constructions for evaluating the infinite nuclear mass Helium ground
state. For our problem, the single bound state of the $\mathrm{H}^{-}$
atom, we shall employ the second basis construction, where exponents
($\alpha_{i}$, for instance) are generated through
\begin{equation}
\alpha_{i}=A_{1}+\{\frac{1}{2}i(i+1)\sqrt{p_{\alpha}}\}(A_{2}-A_{1}),\label{eq:pseudo_ran}
\end{equation}
where $A_{1}$ and $A_{2}$ are the interval bounds, $\{ x\} \equiv x-\lfloor x \rfloor$
denotes the fractional part of $x$, and $p_{\alpha}$ is some prime
number. The same procedure, with different interval bounds and prime
number $p$, is applied to generate $\beta_{i}$ and $\gamma_{i}$
exponents. A complex component will also be added to each of the three
parameters, using the same procedure. In total,  one would have a group
of exponents of size $6N$ \cite{FrolovSmith1995}. In practice, multiple
groups of different sizes must be added to
accurately describe the three-body wavefunction at short (coalescence),
intermediate, and large distances, thereby ensuring improved convergence
of the NR energy.

Unlike hydrogen-like problems, where closed-form expressions for corrections
are readily available, helium-like problems require numerical evaluation
of expectation values, using high-precision energy and wavefunction,
computed following the previously discussed procedure. The effective
Hamiltonian associated with the leading relativistic correction is
expressed as

\begin{equation}
H^{(2)}=H_{\mathrm{si}}^{(2)}+H_{\mathrm{sd}}^{(2)},\label{eq:Hrel}
\end{equation}
where $\mathrm{si}$ and $\mathrm{sd}$ stand for spin-independent and
spin-dependent parts. This correction comes at order $\alpha^{4}mc^{2}$,
and contains recoil effects. The first part is given by the following
sum
\begin{align}
H_{\mathrm{si}}^{(2)} & =H_{\mathrm{kin}.}^{(2)}+H_{\mathrm{Dar}.}^{(2)}+H_{\mathrm{ret}.}^{(2)}.
\end{align}
The first contribution is nothing but the relativistic mass correction,
which can be intuitively derived from Einstein's relativistic energy-momentum
relation,
\begin{align}
H_{\mathrm{kin}.}^{(2)}= & -\frac{\alpha^{2}}{8}\left[\frac{\hat{\mathbf{P}}_{1}^{4}+\hat{\mathbf{P}}_{2}^{4}}{m^{3}}+\frac{\hat{\mathbf{P}}_{0}^{4}}{M^{3}}\right].\label{eq:Rel_Kinetic}
\end{align}
The second contribution is known as the Darwin correction, and in
our case it reads
\begin{equation}
\begin{aligned}
H_{\mathrm{Dar}.}^{(2)}= & -\frac{\pi\alpha^{2}}{m^{2}}\delta(\mathbf{r}_{12})\\
 & +\frac{Z\pi\alpha^{2}}{2}(\frac{1}{m^{2}}+\frac{1}{M^{2}})[\delta(\mathbf{r}_{1})+\delta(\mathbf{r}_{2})].
\end{aligned}\label{eq:Rel_Darwin}
\end{equation}
The factors $1$ in numerators of $1/m^{2}$ and $1/M^{2}$ are initially
given by $(g_{\mathrm{e}}-1)$ and $(g_{\mathrm{N}}-1)$, respectively.
For the electron, the Dirac value $g_{\mathrm{e}}=2$ is used, while its QED corrections are absorbed in higher-order
effective Hamiltonians, such as $H_{\mathrm{SE}}^{(3)}$ and $H_{R_{1}}^{(4)}$,
later reported in Eqs. (\ref{eq:HSE}) and (\ref{eq:highest-order-correc}).
For the nucleus, the large deviation from $g=2$ is absorbed by the
definition of the nuclear radius through the Sachs electric form factor,
as noted in Refs. \cite[section 6.1.1]{Eides2007} and \cite{Jentschura2011}.
Moreover, it is important to note that for spin-$0$ or spin-$1$ nuclei,
such as the alpha-particle or deuteron, the contribution from the
$1/M^{2}$ term must be omitted from consideration, as discussed in
Refs. \cite{Owen1994,PachuckiKarshenboim1995} and \cite[Eq. (12.100)]{JentschuraAdkins2022book}.
This can be compared with the remarks following the hydrogenic Eq. (\ref{eq:EREL}).

The next term is the orbit-orbit correction, which is known from classical
electrodynamics, as a retardation correction to the magnetic interaction
between charged particles \cite[Chapter 12]{Jackson1999}, first derived
by Darwin \cite[page 545]{Darwin1920}
\begin{align}
H_{\mathrm{ret}.}^{(2)} & =+\frac{\alpha^{2}}{2m^{2}}A_{12}-\frac{\alpha^{2}Z}{2mM}\big[A_{10}+A_{20}\big],\label{eq:Ret}\\
A_{ij} & \equiv-r_{ij}^{-1}[\hat{\mathbf{P}}_{i}\cdot\hat{\mathbf{P}}_{j}+r_{ij}^{-2}\mathbf{r}_{ij}\cdot(\mathbf{r}_{ij}\cdot\hat{\mathbf{P}}_{i})\hat{\mathbf{P}}_{j}],\label{eq:Aij}
\end{align}
Concerning the spin-dependent part in Eq. (\ref{eq:Hrel}), for our 
singlet ground state we have $\langle\mathbf{s}_{1}\cdot\mathbf{s}_{2}\rangle=-3/4$
\cite[Eq. (40.9)]{BetheSalpeter1957}, which allows simplifying this term to \cite[Eq. (41.3)]{BetheSalpeter1957}
\begin{equation}
H_{\mathrm{sd}}^{(2)}=\frac{2\pi\alpha^{2}}{m^{2}}\delta(\mathbf{r}_{12});\label{eq:Rel_SD}
\end{equation}
as also noted in Ref. \cite{DRAKE1988}. The complete derivation of
these effective Hamiltonians (relativistic correction) can be found
in Berestetskii \textit{et al.} \cite[\textsection 83]{Berestetskii2012}, for
the interaction between two free spin-$1/2$ particles. The more general case,
where the two spin-$1/2$ particles interact with external fields, is covered in
Bethe and Salpeter \cite[sections 39 and 40]{BetheSalpeter1957} as
well as Jentschura and Adkins \cite[section 12.3]{JentschuraAdkins2022book},
with the latter also addressing anomalous magnetic moments.

Having addressed relativistic corrections for helium-like systems, we
now turn to the first-order QED contributions, whose dominant component,
the self-energy, presents the greatest computational challenge. The
effective Hamiltonian describing the dominant QED contribution, namely
the electron self-energy, for two-electron atoms reads
\begin{equation}
\begin{aligned}H_{\mathrm{SE}}^{(3)} & =\frac{4}{3}\frac{\alpha^{3}Z}{m^{2}}\big(\frac{11}{24}+\frac{3}{8}+\ln(\alpha)^{-2}\\
 & -\ln(k_{0}/\mathrm{Ry})\big)[\delta(\mathbf{r}_{1})+\delta(\mathbf{r}_{2})]\\
 & +\frac{\alpha^{3}}{m^{2}}\big(\frac{17}{3}+\frac{14}{3}\ln\alpha-\frac{20}{3}\mathbf{s}_{1}\cdot\mathbf{s}_{2}\big)\delta(\mathbf{r}_{12})\\
 & -\frac{14}{3}\frac{\alpha^{3}}{m^{2}}Q(\mathbf{r}_{12}),
\end{aligned}\label{eq:HSE}
\end{equation}
and includes the effect of the anomalous magnetic moment of the electron
($3/8$ term), c.f. Eq. (\ref{eq:ESE}). This correction was first
derived by Araki \cite{Araki1957} and Sucher \cite{Sucher1958},
and for more modern derivations, the reader may consult Refs. \cite[section 13.3.2]{JentschuraAdkins2022book}
and \cite{Pachucki1998}. The central technical challenge of this work is the accurate evaluation of the Bethe logarithm $\ln(k_{0}/\mathrm{Ry})$ for negative hydrogen ions, for which we followed the approach of Korobov \cite{Korobov2012}. Because of its critical role in the dominant QED corrections, Appendix \ref{sec:Many-body-Bethe-logarithm} presents its formulation for a general $N$-body system with arbitrary charges and masses, the high-accuracy numerical strategy used in its computation, and possible routes for further improvement. In practice, the Bethe logarithm is evaluated using Eq.~(\ref{eq:BetheTilde}).

As noted above, for the ground state we have $\langle\mathbf{s}_{1}\cdot\mathbf{s}_{2}\rangle=-3/4$, and the expectation value of the $Q(\mathbf{r}_{ij})$-term, with respect to $\tilde{\psi}$ which
solves Eq. (\ref{eq:3body2body}), is given by
\begin{align}
\langle Q(\mathbf{r}_{ij})\rangle & =\lim_{\epsilon\rightarrow0}\int d^{3}r_{1}\int d^{3}r_{2}|\tilde{\psi}(\mathbf{r}_{1},\mathbf{r}_{2})|^{2} \nonumber \\
 &\times\left[\frac{\Theta(r_{ij}-\epsilon/m)}{4\pi r_{ij}^{3}}+\delta(\mathbf{r}_{ij})(\gamma_{E}+\ln\epsilon)\right],\label{eq:Q_term}
\end{align}
where $\gamma_E$ is the Euler-Mascheroni constant, and $\Theta(x)$ is the Heaviside step function.  To this last correction, we shall add the vacuum polarization effect, represented
by the effective Hamiltonian
\begin{align}
H_{\mathrm{VP}}^{(3)}=-\frac{4}{15}\frac{\alpha^{3}}{m^{2}}\big[Z\delta\big(\mathbf{r}_{1}\big)+Z\delta\big(\mathbf{r}_{2}\big)-\delta\big(\mathbf{r}_{12}\big)\big],\label{eq:VP}
\end{align}
c.f. Eq. (\ref{eq:EVP}), to form the total Hamiltonian
\begin{align}
H^{(3)}=H_{\mathrm{SE}}^{(3)}+H_{\mathrm{VP}}^{(3)}.
\end{align}
The first two terms of Eq. (\ref{eq:VP}) account for the vacuum polarization
effect that screens the interaction between the nucleus and each electron,
while the remaining one describes the electron-electron vacuum polarization
screening. 

The first-order recoil correction to this two-electron Lamb shift
was discussed in detail by Pachucki and Sapirstein in Refs. \cite{Pachucki1998,PachuckiSapirstein2000}.
We retain only terms that do not vanish for singlet states. Taking these corrections into account yields the following  Lamb shift
expression, given by
\begin{align}
\langle H^{(3)}\rangle_{\infty}+E_{A}^{(3)}+E_{B}^{(3)},
\end{align}
where the last two terms account for the first-order recoil correction.
The first term reads
\begin{align}
E_{A}^{(3)} & = \frac{\alpha^{3}Z^{2}}{mM}\bigg[\frac{2}{3}\ln(\alpha^{-1})+\frac{62}{9} \nonumber \\
 & -\frac{8}{3}\ln(k_{0}^{\infty}/\mathrm{Ry})\bigg]\langle\delta(\mathbf{r}_{1})+\delta(\mathbf{r}_{2})\rangle_{\infty} \nonumber \\
 & -\frac{14}{3}\frac{\alpha^{3}Z^{2}}{Mm}\langle Q(\mathbf{r}_{1})+Q(\mathbf{r}_{2})\rangle_{\infty},
\end{align}
and accounts for recoil correction at the operator level. Here, the notation $\langle \cdot \rangle_{\infty}$ indicates expectation values evaluated with the infinite-nuclear-mass wavefunction, and $\ln(k_{0}^{\infty}/\mathrm{Ry})$ represents the infinite nuclear mass Bethe logarithm, computed using Eqs. (\ref{eq:BetheTilde}), with an infinite-nuclear-mass energy, wavefunction, Hamiltonian, and $\tilde{\mathbf{J}}$ (see Appendix \ref{sec:Many-body-Bethe-logarithm}). The remaining contribution comes from the recoil correction at the wavefunction level, in $\langle H^{(3)}\rangle$. This last correction reads
\begin{align}E_{B}^{(3)} & =\frac{4}{3}\frac{\alpha^{3}Z}{m^{2}}\bigg[\ln(\alpha)^{-2}\nonumber\\
 &  -\ln(k_{0}^{\infty}/\mathrm{Ry})+\frac{19}{30}\bigg]\langle\delta(\mathbf{r}_{1})+\delta(\mathbf{r}_{2})\rangle_{\mathrm{rec}}\nonumber\\
 & +\frac{\alpha^{3}}{m^{2}}\bigg[\frac{14}{3}\ln\alpha+\frac{164}{15}\bigg]\langle\delta(\mathbf{r}_{12})\rangle_{\mathrm{rec}}\nonumber\\
 & -\frac{14}{3}\frac{\alpha^{3}}{m^{2}}\langle Q(\mathbf{r}_{12})\rangle_{\mathrm{rec}}\nonumber\\
 & -\frac{4}{3}\frac{\alpha^{3}Z}{m^{2}}\ln(k_{0}^{\mathrm{rec}}/\mathrm{Ry})\langle\delta(\mathbf{r}_{1})+\delta(\mathbf{r}_{2})\rangle_{\infty},
\end{align}
following Refs. \cite{Pachucki1998,PachuckiSapirstein2000}. In this
equation, $\langle \cdot \rangle_{\mathrm{rec}}$ and $\ln(k_{0}^{\mathrm{rec}}/\mathrm{Ry})$
represent the first-order recoil correction to the expectation values
and the Bethe logarithm, respectively. On the other hand, second-order
recoil corrections $(m/M)^{2}$ for helium-like problems were very
recently considered by Pachucki \textit{et al.} \cite{pachucki2025}. 
Including this effect, and neglecting the spin-dependent contributions, which enter only at order $(m/M)^3$, one can write the total QED energy shift, up
to the second-order recoil correction as \cite[Eq. (9)]{pachucki2025}
\begin{widetext}
\begin{align}E_{\mathrm{total}}^{(3)} & =-\frac{4}{3}\alpha^{3}Z\left(\frac{1}{m}+\frac{Z}{M}\right)^{2}\ln(k_{0}^{M/m}/\mathrm{Ry})\langle\delta(\mathbf{r}_{1})+\delta(\mathbf{r}_{2})\rangle_{M/m}\nonumber\\
 & +\frac{\alpha^{3}}{m^{2}}\bigg\{\frac{4}{3}Z\left(\ln(\alpha^{-2})+\frac{19}{30}\right)+Z^{2}\frac{m}{M}\left(\frac{1}{3}\ln(\alpha^{-2})+\frac{62}{9}\right)+\frac{4}{3}Z^{3}\left(\frac{m}{M}\right)^{2}\ln\left(\frac{M}{m\alpha^{2}}\right)\bigg\}\langle\delta(\mathbf{r}_{1})+\delta(\mathbf{r}_{2})\rangle_{M/m}\nonumber\\
 & +\frac{\alpha^{3}}{m^{2}}\left(\frac{14}{3}\ln(\alpha)+\frac{164}{15}\right)\langle\delta(\mathbf{r}_{12})\rangle_{M/m}-\frac{14}{3}\frac{\alpha^{3}}{m^{2}}\langle Q(\mathbf{r}_{12})\rangle_{M/m}-\frac{14}{3}\frac{\alpha^{3}Z^{2}}{Mm}\langle Q(\mathbf{r}_{1})+Q(\mathbf{r}_{2})\rangle_{M/m}.
 \label{eq:E3tot}
\end{align}
\end{widetext}

We shall evaluate this expression in our calculations of the leading-order QED correction to the negative hydrogen ions.
In this expression, the $\langle \cdot \rangle_{M/m}$ symbols indicate that all expectation values are taken with respect to the finite-nuclear-mass three-body wavefunction. The Bethe logarithm, entering the first line, $\ln(k_{0}^{M/m}/\mathrm{Ry}) = \ln(k_{0}/\mathrm{Ry})$ contains all-order recoil effects. 
Two points are worth noting regarding the second-order recoil contribution
of order $(m/M)^{2}$ to the leading-order QED nuclear self-energy
correction (see the hydrogenic counterpart in Eq. (\ref{eq:NSE})).
First, the nonlogarithmic term is not known for nuclei with spin $I\neq1/2$
\cite{Puchalski2019} (e.g. the deuteron case) but is numerically
negligible, at the present level of precision. Second, for spin-$1/2$
nuclei such as the proton, this contribution has been evaluated in
Ref. \cite{Puchalski2012H2} in the context of the four-body $\mathrm{H}_{2}$
problem, where it was absorbed into the definition of the proton charge
radius ($r_{\mathrm{p}}$) following the proposal of Pachucki \cite{Pachucki1995}.
In the present work, all second-order recoil effects are well below
our overall uncertainty, so the omission of this nonlogarithmic term
has no impact on the calculated energies of the negative hydrogen
ions.

We next consider the contribution $E^{(4)}\sim\alpha^{6}mc^{2}$,
comprising second-order relativistic and QED effects, previously considered
in Refs. \cite{Pachucki2006,KorobovYelkhovsky2001,Yelkhovsky2001}.
As in the helium case, preliminary estimates show that this correction
is overwhelmingly dominated by the effective Hamiltonian \cite[Eq. (2.56)]{Pachucki2006}\cite[Eqs. (33 and 34)]{Yelkhovsky2001}
\begin{equation}
\begin{aligned} H_{R_{1}}^{(4)} &\!= \! \frac{\alpha^{4}}{m^{2}}Z^{2}\left[\frac{427}{96}-2\ln(2)\right]\pi\big[\delta(\mathbf{r}_{1})+\delta(\mathbf{r}_{2})\big]\\
 & \!+\! \frac{\alpha^{4}}{m^{2}}\left[\frac{6\zeta(3)}{\pi^{2}}\!-\!\frac{697}{27\pi^{2}}\!-\!8\ln(2)\!+\!\frac{1099}{72}\right]\!\pi\delta(\mathbf{r}_{12}),\label{eq:highest-order-correc}
\end{aligned}
\end{equation}
c.f. Eq. (\ref{eq:E4r1}), originating from one-loop radiative corrections.  At the present level
of precision, recoil effects are negligible. The sum of the remaining components
of $H^{(4)}$ is numerically insignificant (see \cite{Pachucki2006}), and higher-order corrections
($H^{(5)}$ and beyond) are expected to be minor. We therefore assign, very conservatively, an ultimate uncertainty equal to one third of the shift induced by $H_{R_{1}}^{(4)}$, which we take as the ultimate
error on the ground-state energy of the anion and, consequently, on
its photodetachment energy.

We conclude our corrections by accounting for the finite-nuclear-size (FNS) effect in the two-electron atomic problem, described by the effective Hamiltonian
\begin{equation}
H^{\mathrm{FNS}}=\frac{2\pi}{3}\frac{Z\alpha^{2}}{m^{2}}(r_{\mathrm{N}}/\lambdabar)^{2}\big[\delta(\mathbf{r}_{1})+\delta(\mathbf{r}_{2})\big],\label{eq:FNS}
\end{equation}
c.f. Eq. (\ref{eq:EFNS}). We now turn to the numerical evaluation of the ground-state energies of hydrogen isotopes and their corresponding negative ions.

\section{Numerical evaluations}

\label{sec:Numerical-evaluations}

In our numerical evaluations, we shall employ the CODATA 2018 recommended
physical constants \cite{Tiesinga2021RevModPhys.93.025010}
\begin{equation}
\begin{aligned}M_{\mathrm{p}}/m & =1836.15267343\\
M_{\mathrm{d}}/m & =3670.48296788\\
M_{\mathrm{t}}/m & =5496.92153573\\
\alpha^{-1} & =137.035999084\\
a_{0} & =5.29177210903\times10^{-11}\,\mathrm{m}\\
\lambdabar & =3.8615926796\times10^{-13}\,\mathrm{m}\\
g_{\mathrm{e}} & =2.00231930436256\\
g_{\mathrm{p}} & =5.5856946893\\
g_{\mathrm{d}} & =0.8574382338\\
g_{\mathrm{t}} & =5.957924931\\
r_{\mathrm{p}} & =0.8414\,\mathrm{fm}\\
r_{\mathrm{d}} & =2.12799\,\mathrm{fm}.
\end{aligned}
\end{equation}
We adopt the experimentally determined triton charge radius $r_{\mathrm{t}}=1.755\,\mathrm{fm}$
from Ref. \cite{AMROUN1994596}. The precision of the present results
is unaffected by the uncertainties of the fundamental constants. Moreover,
the updated CODATA 2022 values \cite{CODATA20222025} differ from
CODATA 2018 by amounts well within our numerical error bars, leaving
the reported results unchanged.

\subsection{Neutral hydrogen isotopes}

\label{subsec:Neutral-hydrogen-isotopes}

We shall first start with the hydrogen problem, discussed in Section
\ref{subsec:Hydrogenic-section}. We evaluated all corrections given
in Eqs. (\ref{eq:ENR}, \ref{eq:EREL}, \ref{eq:ESE}, \ref{eq:EVP},
\ref{eq:EFNS}, and \ref{eq:EHFS}), and reported results in Table
\ref{tab:hydrogen}. It is worth noting that the reported digits are
only limited by our knowledge of above fundamental constants. The
Bethe logarithm of the hydrogen ground-state is $\ln(k_{0}/\mathrm{Ry})=2.984\,128\,556...$,
computed in Ref. \cite{DrakeSwainson1990}. A closed analytical expression
for this Bethe logarithm was provided in Ref. \cite[Eq. (4.337)]{JentschuraAdkins2022book}.

\begin{table}[htbp]
\centering
\begin{tabular}{|l|l|l|l|}
\cline{2-4}
\multicolumn{1}{l|}{} & $^{1}\mathrm{H}(F=0)$ & $^{2}\mathrm{H}(F=1/2)$ & $^{3}\mathrm{H}(F=0)$\tabularnewline
\hline 
$E^{(0)}$ & $-0.499727839712$ & $-0.499863815247$ & $-0.499909056541$\tabularnewline
\hline 
$E^{(2)}$ & $-0.000006656415$ & $-0.000006656420$ & $-0.000006656419$\tabularnewline
\hline 
$E^{(3)}$ & $+0.000001233673$ & $+0.000001234452$ & $+0.000001234711$\tabularnewline
\hline 
$E^{(4)}$ & $+0.000000008645$ & $+0.000000008652$ & $+0.000000008655$\tabularnewline
\hline 
$E^{\mathrm{FNS}}$ & $+0.000000000168$ & $+0.000000001077$ & $+0.000000000733$\tabularnewline
\hline 
Total & $-0.49973325364$ & $-0.49986922749$ & $-0.49991446886$\tabularnewline
\hline 
\hline 
$E^{\mathrm{HF}}$ & $-0.00000016192$ & $-0.00000003317$ & $-0.00000017290$\tabularnewline
\hline 
Total & $-0.49973341556$ & $-0.49986926065$ & $-0.49991464176$\tabularnewline
\hline 
\end{tabular}

\caption{\label{tab:hydrogen}Leading contributions to the hyperfine ground state of neutral hydrogen isotopes, in atomic units.}
\end{table}

In the following, we shall consider the numerical calculations done
on the corresponding negative hydrogen isotopes.

\subsection{Negative hydrogen isotopes}

\label{subsec:Negative-hydrogen-isotopes}

Concerning the numerical determination of the nonrelativistic ground-state energy, we have performed a set of calculations on infinite-nuclear-mass $^{\infty}\mathrm{H}^{-}$ and on the finite-nuclear-mass negative isotopes $^{1}\mathrm{H}^{-}$, $^{2}\mathrm{H}^{-}$, and $^{3}\mathrm{H}^{-}$. For each system, calculations were carried out with systematically increasing basis-set sizes and extrapolated to the infinite-basis-set limit by least-squares fitting the data to $E(N)=a+b/N^{c}$, yielding the parameters $a$, $b$, and $c$. This procedure uses an empirical power-law form reflecting the observed asymptotic convergence of the variational energy with basis size. The nonlinear exponents entering the basis were optimized to minimize the variational energy, and calculations were performed for basis-set sizes in the range $N=3500$--$4500$ prior to extrapolation to the $N\to\infty$ limit. The infinite-basis-set limit is then given by the parameter $a$, with its associated standard error providing the corresponding uncertainty.

We report the resulting infinite-nuclear-mass ground-state energy and compare it with previously reported values in Table~\ref{tab:inf_energy}. TW stands for ``This Work''. Our computed ground-state energy is slightly higher than the most accurate published values (Nakashima and Nakatsuji~\cite{NakashimaNakatsuji2007}; Aznabaev \textit{et al.}~\cite{AznabaevBekbaevKorobov2018} ; Korobov and Bu\v sa~\cite{Korobov2025}), and our small quoted uncertainty is likely underestimated. Achieving a lower variational energy and a more reliable uncertainty estimate would require adding larger exponents to the basis to improve convergence as $N\rightarrow\infty$. This lies beyond the scope of the present work, as it does not affect the targeted precision of the photodetachment energy. We next report the ground-state energies for the three isotopes in Table \ref{tab:fin_energy} and compare them with those of Petrimoulx \textit{et al.} \cite{PetrimoulxBondyEneSatiDrake2025} who use the same mass ratios (CODATA2018), and Aznabaev \textit{et al.} \cite{AznabaevBekbaevKorobov2018}. These comparisons indicate that our calculations reach a high level of numerical precision, with 27 digits being converged. 

\begin{table*}[t]
\centering
\begin{tabular}{|c|c|l|l|}
\hline 
Author & Year & Energy (in a.u.) & Ref.\tabularnewline
\hline 
\hline 
Bethe & 1929 & $-0.5253$ & \cite{Bethe1929}\tabularnewline
\hline 
Hylleraas & 1930 & $-0.5264$ & \cite{Hylleraas1930}\tabularnewline
\hline 
Henrich & 1944 & $-0.52756$ & \cite{Henrich1944}\tabularnewline
\hline 
Hylleraas and Midtdal & 1956 & $-0.52772\;6$ & \cite{HylleraasMidtdal1956}\tabularnewline
\hline 
Pekeris & 1958 & $-0.52775\;0962$ & \cite{Pekeris1958}\tabularnewline
\hline 
Pekeris & 1962 & $-0.52775\;1014$ & \cite{Pekeris1962}\tabularnewline
\hline 
Frankowski and Pekeris & 1966 & $-0.52775\;10163\;8$ & \cite{FrankowskiPekeris1966}\tabularnewline
\hline 
Frolov & 1987 & $-0.52775\;10165\;0$ & \cite{frolov1987}\tabularnewline
\hline 
Drake & 1988 & $-0.52775\;10165\;44306(85)$ & \cite{DRAKE1988}\tabularnewline
\hline 
Drake \textit{et al.} & 2002 & $-0.52775\;10165\;44377\;196613(22)$ & \cite{DrakeCassarNistor2002}\tabularnewline
\hline 
Frolov & 2006 & $-0.52775\;10165\;44377\;19658\;9759$ & \cite{Frolov2006}\tabularnewline
\hline 
Nakashima and Nakatsuji & 2007 & $-0.52775\;10165\;44377\;19659\;08145\;66747\;511$ & \cite{NakashimaNakatsuji2007}\tabularnewline
\hline 
Aznabaev \textit{et al.} & 2018 & $-0.52775\;10165\;44377\;19659\;08145\;66747\;5776$ & \cite{AznabaevBekbaevKorobov2018}\tabularnewline
\hline 
Petrimoulx \textit{et al.} & 2025 & $-0.52775\;10165\;44377\;19659\;08145\;03(5)$ & \cite{PetrimoulxBondyEneSatiDrake2025}\tabularnewline
\hline
Korobov and Bu\v sa & 2025 & $-0.52775\;10165\;44377\;19659\;08145\;66747\;57779\;67$ & \cite{Korobov2025}\tabularnewline
\hline 
Salman and Karr & 2026 & $-0.52775\;10165\;44377\;19659\;08145\;6639(14)$ & TW\tabularnewline
\hline 
\end{tabular}
\caption{\label{tab:inf_energy}Theoretical predictions of the infinite
proton mass $\mathrm{H}^{-}$ nonrelativistic ground-state energy (in $\mathrm{a.u.}$).}
\end{table*}

\begin{table}[htbp]
\centering
\begin{tabular}{|r|l|l|}
\cline{2-3}
\multicolumn{1}{r|}{} & Energy (in a.u.) & Ref.\tabularnewline
\hline 
$^{1}\mathrm{H}^{-}$ & $-0.52744\;58811\;09600\;16257\;27344\;71(5)$ & \cite{PetrimoulxBondyEneSatiDrake2025}\tabularnewline
\cline{2-3}
 & $-0.52744\;58811\;09600\;16257\;27346\;2055(14)$ & TW\tabularnewline
\hline 
$^{2}\mathrm{H}^{-}$ & $-0.52759\;83246\;84559\;87412\;08883\;15(5)$ & \cite{PetrimoulxBondyEneSatiDrake2025}\tabularnewline
\cline{2-3}
 & $-0.52759\;83246\;84559\;87412\;08883\;5744(14)$ & TW\tabularnewline
\hline 
$^{3}\mathrm{H}^{-}$ & $-0.52764\;90482\;02178\;91232\;43743\;46(5)$ & \cite{PetrimoulxBondyEneSatiDrake2025}\tabularnewline
\cline{2-3}
 & $-0.52764\;90482\;02178\;91232\;43744\;1259(14)$ & TW\tabularnewline
\hline 
$^{\infty}\mathrm{H}^{-}$ & $-0.52775\;10165\;44377\;19659\;08145\;66747\;577\ldots$ & \cite{Korobov2025}\tabularnewline
\cline{2-3}
 & $-0.52775\;10165\;44377\;19659\;08145\;6639(14)$ & TW\tabularnewline
\hline 
\end{tabular} 
\caption{\label{tab:fin_energy}Theoretical predictions of the nonrelativistic
ground-state energies of negative hydrogen ions.}
\end{table}

Next, we evaluate expectation values of the one- and two-body operators entering the corrections discussed in Sec.~\ref{subsec:Helium-like-section}.  These expectation values do not obey a variational principle. However, upon increasing the basis-set size $N$, they stabilize beyond a threshold value, while exhibiting fluctuations due to the finite arithmetic precision. We therefore take the expectation values and their uncertainties as the mean and standard deviation, respectively, of the results obtained within this stability window. The extent of this window depends on the operator under consideration; overall, basis sizes ranging from a few thousand up to $\sim 10^{4}$ functions were required.

For our singlet ground state, the wavefunction is symmetric under particle exchange $\mathbf{r}_1 \leftrightarrow \mathbf{r}_2$, implying $\langle \delta(\mathbf{r}_1) \rangle = \langle \delta(\mathbf{r}_2) \rangle$ and analogous relations for the remaining expectation values. All results reported here surpass the precision of previously available data. Expectation values of the $\delta(\mathbf{r}_1)$ and $\delta(\mathbf{r}_{12})$ operators are listed in Table~\ref{tab:delta_expectation}, those of the relativistic operators $\mathbf{P}_i^{4}$ and $A_{ij}$ in Tables~\ref{tab:P4_expectation} and \ref{tab:A_expectation}, respectively, while the $Q$-term expectation values entering the self-energy correction are given in Table~\ref{tab:Q_expectation}.

\begin{table}[htbp]
\centering
\begin{tabular}{|r|l|l|l|}
\cline{2-4}
\multicolumn{1}{r|}{} & $\langle \delta(\mathbf{r}_{1})\rangle\times10$ & $\langle\delta(\mathbf{r}_{12})\rangle\times10^{3}$ & Ref.\tabularnewline
\hline 
$^{1}\mathrm{H}^{-}$ & $1.642799441188473(14)$ & $2.73157279257130(16)$ & TW\tabularnewline
\hline 
$^{2}\mathrm{H}^{-}$ & $1.644162633549041(12)$ & $2.73477836046327(17)$ & TW\tabularnewline
\hline 
$^{3}\mathrm{H}^{-}$ & $1.644616368107743(15)$ & $2.73584562773331(18)$ & TW\tabularnewline
\hline 
$^{\infty}\mathrm{H}^{-}$ & $1.6455287286(3)$ & $2.7379923(30)$ & \cite{Drake1996}\tabularnewline
\cline{2-4}
 & $1.6455287284713$ & $2.7379921262294$ & \cite{Frolov2007}\tabularnewline
\cline{2-4}
 & $1.645528728472080(12)$ & $2.73799212612099(16)$ & TW\tabularnewline
\hline 
\end{tabular}
\caption{\label{tab:delta_expectation}Expectation
values of $\delta(\mathbf{r}_{1})$ and $\delta(\mathbf{r}_{12})$,
in atomic units.}
\end{table}

\begin{table}[htbp]
\centering
\begin{tabular}{|r|l|l|l|}
\cline{2-4}
\multicolumn{1}{r|}{} & $\langle\hat{\mathbf{P}}_{0}^{4}\rangle$ & $\langle\hat{\mathbf{P}}_{1}^{4}\rangle$ & Ref.\tabularnewline
\hline 
$^{1}\mathrm{H}^{-}$ & $5.634006989831(42)$ & $2.457148087386454(43)$ & TW\tabularnewline
\hline 
$^{2}\mathrm{H}^{-}$ & $5.640645098676(42)$ & $2.459850106386986(41)$ & TW\tabularnewline
\hline 
$^{3}\mathrm{H}^{-}$ & $5.642855170010(42)$ & $2.460749627702028(41)$ & TW\tabularnewline
\hline 
$^{\infty}\mathrm{H}^{-}$ & $---------$ & $2.46255856(12)$ & \cite{DRAKE1988}\tabularnewline
\cline{2-4}
 & $---------$ & $2.462558614(3)$ & \cite{Drake1996}\tabularnewline
\cline{2-4}
 & $5.647300038761(43)$ & $2.462558612368980(43)$ & TW\tabularnewline
\hline 
\end{tabular}
\caption{\label{tab:P4_expectation}Expectation values of $\mathbf{P}_{0}^{4}$
and $\mathbf{P}_{1}^{4}$, entering Eq. (\ref{eq:Rel_Kinetic}),
in atomic units.}
\end{table}

\begin{table*}[t]
\centering
\begin{tabular}{|r|l|l|l|}
\cline{2-4}
\multicolumn{1}{r|}{} & $\langle A_{10}\rangle $ & $\langle A_{12}\rangle \times10^{2}$ & Ref.\tabularnewline
\hline 
$^{1}\mathrm{H}^{-}$ & $1.012806087936993052(45)$ & $-1.7685772444054362(25)$ &  TW\tabularnewline
\hline 
$^{2}\mathrm{H}^{-}$ & $1.013683422743481543(59)$ & $-1.7717860617518059(25)$ & TW\tabularnewline
\hline 
$^{3}\mathrm{H}^{-}$ & $1.013975456290267090(43)$ & $-1.7728547058132548(25)$ & TW\tabularnewline
\hline 
$^{\infty}\mathrm{H}^{-}$ & $-----------$ & $-1.77500446(4)$ & \cite{DRAKE1988}\tabularnewline
\cline{2-4}
 & $-----------$ & $-1.775004420(2)$ & \cite{Drake1996}\tabularnewline
\cline{2-4}
 &  $1.014562694548989662(45)$ &$-1.7750044190985449(24)$ & TW\tabularnewline
\hline 
\end{tabular}
\caption{\label{tab:A_expectation}Expectation values of $A_{12}$
and $A_{10}$, defined in Eq. (\ref{eq:Aij}), in atomic
units.}
\end{table*}

\begin{table}[htbp]
\centering
\begin{tabular}{|r|l|l|l|}
\cline{2-4}
\multicolumn{1}{r|}{} & $\langle Q(\mathbf{r}_{1})\rangle \times10$ & $\langle Q(\mathbf{r}_{12})\rangle \times10^{3}$ & Ref.\tabularnewline
\hline 
$^{1}\mathrm{H}^{-}$ & $-1.085748458405272(96)$ & $7.8421419495946(20)$ & TW\tabularnewline
\hline 
$^{2}\mathrm{H}^{-}$ & $-1.087088024331203(97)$ & $7.8487892087935(15)$ & TW\tabularnewline
\hline 
$^{3}\mathrm{H}^{-}$ & $-1.087533970896739(99)$ & $7.8510018565553(15)$ & TW\tabularnewline
\hline 
$^{\infty}\mathrm{H}^{-}$ & $-----------$ & $7.8554(1)$ &\cite{DRAKE1988}\tabularnewline
\cline{2-4}
& $-1.08843078615236$ & $7.8554511967292$  & \cite{Frolov2013}\tabularnewline
\cline{2-4}
  & $-1.088430786142999(97)$ & $7.8554511965462(41)$ & TW\tabularnewline
\hline 
\end{tabular}
\caption{$Q(\mathbf{r}_{ij})$ expectation values, defined in Eq. (\ref{eq:Q_term}), in atomic units.}\label{tab:Q_expectation}
\end{table}

The most challenging quantity to evaluate is the Bethe logarithm (see Sec.~\ref{subsec:Helium-like-section}). Following the formalism outlined in Appendix~\ref{sec:Many-body-Bethe-logarithm}, we compute the (three-body) Bethe logarithm $\ln(k_{0}/\mathrm{Ry})$, defined in Eq.~(\ref{eq:BetheTilde}), for the three negative isotopes and for the infinite-nuclear-mass limit. The Bethe logarithm is evaluated using reference-state basis sets with sizes $N \in [4500,5500]$  and intermediate-state basis sets with sizes $N \in [4000,8000]$, the latter used to represent the eigenstates $|n\rangle$ entering the spectral representation of the resolvent in Eq.~(\ref{eq:resolvent}) and including large exponents to capture high-energy contributions (see \cite[Sec. III.A]{Korobov2012}). Our results are reported in Table~\ref{tab:Bethe_log_results} and compared with available values from the literature. 

The dominant uncertainty arises from the extrapolation parameters $A_i$, $B_i$, and $C_i$ ($i \ge 4$) (see Eq.~(\ref{eq:Jrem})). The $\delta$-function expectation values in $\cal{D}$ (Eq.~(\ref{eq:D_sum})) and the low-energy integral (first term in Eq.~(\ref{eq:BetheTilde})) contribute at the same level, about one order of magnitude smaller. All other contributions are negligible: those from $C_{3}$ (Eq.~(\ref{eq:A3B3C3})) and $\tilde{\mathbf{J}}^{2}$ (second term in Eq.~(\ref{eq:BetheTilde})) are smaller by roughly two and eight orders of magnitude, respectively. The achieved precision on the Bethe logarithm largely exceeds present and foreseeable requirements, as higher-order corrections dominate the theoretical uncertainty. While further
improvement of the expectation values is straightforward, reducing
the dominant extrapolation error requires a more substantial refinement
of the fitting scheme (see Appendix \ref{sec:Many-body-Bethe-logarithm}).

\begin{table}[htbp]
\centering
\begin{tabular}{|r|c|l|l|}
\cline{2-4}
\multicolumn{1}{r|}{} & date & $\ln(k_{0}/\mathrm{Ry})$ & Ref.\tabularnewline
\hline 
$^{1}\mathrm{H}^{-}$ & 2000 & $2.9924809(1)$ & \cite{DrakeGoldman2000}\tabularnewline
\cline{2-4}
 & 2005 & $2.99248087$ & \cite{FROLOV2005}\tabularnewline
\cline{2-4}
 & 2015 & $2.992480900$ & \cite{Frolov2015}\tabularnewline
\cline{2-4}
 & 2026 & $2.99248114559933(54)$ & TW\tabularnewline
\hline 
$^{2}\mathrm{H}^{-}$ & 2000 & $2.9927425(1)$ & \cite{DrakeGoldman2000}\tabularnewline
\cline{2-4}
 & 2005 & $2.99274247$ & \cite{FROLOV2005}\tabularnewline
\cline{2-4}
 & 2015 & $2.992742490$ & \cite{Frolov2015}\tabularnewline
\cline{2-4}
 & 2026 & $2.99274274825907(54)$ & TW\tabularnewline
\hline 
$^{3}\mathrm{H}^{-}$ & 2000 & $2.9928295(1)$ & \cite{DrakeGoldman2000}\tabularnewline
\cline{2-4}
 & 2005 & $2.99282950$ & \cite{FROLOV2005}\tabularnewline
\cline{2-4}
 & 2015 & $2.992829511$ & \cite{Frolov2015}\tabularnewline
\cline{2-4}
 & 2026 & $2.99282976946737(54)$ & TW\tabularnewline
\hline 
$^{\infty}\mathrm{H}^{-}$ & 1970 & $2.984$ & \cite{AASHAMAR1970}\tabularnewline
\cline{2-4}
 & 1984 & $2.9718$ & \cite{GoldmanDrake1984}\tabularnewline
\cline{2-4}
 & 1988 & $2.9718$ & \cite{DRAKE1988}\tabularnewline
\cline{2-4}
 & 2000 & $2.99297(5)$ & \cite{Baker2000bethe}\tabularnewline
\cline{2-4}
 & 2000 & $2.9930044(1)$ & \cite{DrakeGoldman2000}\tabularnewline
\cline{2-4}
 & 2005 & $2.99300440$ & \cite{FROLOV2005}\tabularnewline
\cline{2-4}
 & 2015 & $2.993004415$ & \cite{Frolov2015}\tabularnewline
\cline{2-4}
 & 2026 & $2.99300467073613(54)$ & TW\tabularnewline
\hline 
\end{tabular}\caption{Values of the Bethe logarithm, $\ln(k_{0}/\mathrm{Ry})$, given in
Eq. (\ref{eq:BetheTilde}), and discussed in Sec. \ref{sec:Many-body-Bethe-logarithm}.}\label{tab:Bethe_log_results}
\end{table}

\subsection{Photodetachment energies}

\label{subsec:Photodetachment-energy-of}

To conclude, we present our results for the photodetachment energy
$E_{\mathrm{PD}}=E(\mathrm{A})-E(\mathrm{A}^{-})$. Here, the neutral atom
$\mathrm{A}$ stands for $^{1}\mathrm{H}$, $^{2}\mathrm{H}$, or $^{3}\mathrm{H}$.
The individual contributions to $E_{\mathrm{PD}}$ are given by $E_{\mathrm{PD}}^{(i)}=E^{(i)}(\mathrm{A})-E^{(i)}(\mathrm{A}^{-})$,
with $i=0,2,3,4$, in addition to the FNS, as well HF correction.
The hydrogenic contributions $E^{(i)}(\mathrm{A})$ are discussed in
Section \ref{subsec:Hydrogenic-section}, and the corresponding helium-like
terms $E^{(i)}(\mathrm{A}^{-})$ in Section \ref{subsec:Helium-like-section}.
Our results are reported
in Table \ref{tab:PDEcontributions}. 
For the $E_{\mathrm{PD}}^{(4)}$ contribution, we retain only the hydrogenic term $E^{(4)}_{R_1}$ of Eq.~(\ref{eq:E4r1}), for consistency with the negative-ion treatment in Eq.~(\ref{eq:highest-order-correc}). This choice is further justified by the strong cancellation among the remaining terms, as shown in Table~II of Ref.~\cite{Pachucki2006}. We recall that the conservative
error bar of $3.1\times10^{-9}\,\mathrm{a.u.}$ is set to be one third
of $E^{(4)}(\mathrm{A}^{-})$, which we take
to cover the impact of higher-order corrections $E_{\mathrm{PD}}^{(5)}$
as well as the residual components of $E^{(4)}(\mathrm{A}^{-})$ not
included in the present work (see Sec. \ref{subsec:Helium-like-section}).

\begin{table*}[t]
\centering
\begin{tabular}{|l|l|l|l|}
\cline{2-4}
\multicolumn{1}{c|}{} & $^{1}\mathrm{H}^{-}$ & $^{2}\mathrm{H}^{-}$ & $^{3}\mathrm{H}^{-}$\tabularnewline
\hline 
$E_{\mathrm{PD}}^{(0)}$ & $+0.027718041397$ & $+0.027734509437$ & $+0.027739991661$\tabularnewline
\hline 
$E_{\mathrm{PD}}^{(2)}$ & $-0.000001384574$ & $-0.000001385753$ & $-0.000001386146$\tabularnewline
\hline 
$E_{\mathrm{PD}}^{(3)}$ & $-0.000000013293$ & $-0.000000013315$ & $-0.000000013323$\tabularnewline
\hline 
$E_{\mathrm{PD}}^{(4)}$ & $-0.0000000005(31)$ & $-0.0000000005(31)$ & $-0.0000000005(31)$\tabularnewline
\hline 
$E_{\mathrm{PD}}^{\mathrm{FNS}}$ & $-0.000000000006$ & $-0.000000000037$ & $-0.000000000025$\tabularnewline
\hline 
\hline 
$E_{\mathrm{PD}}$ & $+0.0277166430(31)$ & $+0.0277331098(31)$ & $+0.0277385917(31)$\tabularnewline
\hline 
\hline 
$E^{\mathrm{HF}}$ & $-0.000000161917$ & $-0.000000033167$ & $-0.000000172895$\tabularnewline
\hline 
$E_{\mathrm{PD}}^{\mathrm{HF}}$ & $+0.0277164811(31)$ & $+0.0277330767(31)$ & $+0.0277384188(31)$\tabularnewline
\hline 
\end{tabular}
    \caption{Contributions to the photodetachment energy of $^{1}\mathrm{H}^{-}$,
$^{2}\mathrm{H}^{-}$, and $^{3}\mathrm{H}^{-}$, reported in atomic units.}\label{tab:PDEcontributions}
\end{table*}

To the best of our knowledge, no experimental measurement exists for
the photodetachment energy of the negative tritium ion. For the negative
hydrogen and deuterium ions, a comparison between experimental measurements
and theoretical predictions is given in Table \ref{tab:Final-PD-comparison}.
We note that Beyer and Merkt \cite{BeyerMerkt2018} reported two different
values for the $^{2}\mathrm{H}^{-}$ photodetachment energy, obtained
using two distinct analysis methods: one based on the thermochemical
cycle {[}their Eq. (1){]} and the other on the threshold-interval
relation {[}their Eq. (3){]}, the latter offering higher accuracy
owing to reduced sensitivity to field-induced shifts, as claimed by
the authors.

\begin{table}[htbp]
\centering
\begin{tabular}{|l|l|l|l|l|}
\cline{2-5}
\multicolumn{1}{l|}{} & $^{1}\mathrm{H}^{-}$ & $^{2}\mathrm{H}^{-}$ & Ref.    & Year\tabularnewline
\hline 
Expe. & $6082.99(15)$    & $6086.2(6)$      & \cite{Lykke1991PhysRevA.43.6104} & 1991\tabularnewline
\cline{2-5} 
      & $6082.8(7)$      & $-------$        & \cite{Oliver_Harms_1997}         & 1997\tabularnewline
\cline{2-5}
      & $6083.6(5)$      & $6087.2(5)$      & \cite{BeyerMerkt2018}            & 2018\tabularnewline
\cline{2-5}
      & $-------$        & $6086.81(27)$    & \cite{BeyerMerkt2018}            & 2018\tabularnewline
\hline 
Theo. & $6083.063877$    & $-------$        & \cite{DRAKE1988}$\ensuremath{^{*}}$ & 1988\tabularnewline
\cline{2-5}
      & $6083.06447(68)$ & $6086.70679(68)$ & TW & 2026\tabularnewline
\hline 
\end{tabular}\caption{Comparison between experimental measurements and theoretical predictions of the photodetachment energies of ${}^{1}\mathrm{H}^{-}$ to ${}^{1}\mathrm{H}(F=0)$ and ${}^{2}\mathrm{H}^{-}$ to ${}^{2}\mathrm{H}(F=1/2)$ (in $\mathrm{cm}^{-1}$). $({^{*}})$ hyperfine correction ($F=0$) added for ${}^{1}\mathrm{H}^{-}$ photodetachement of Ref. \cite{DRAKE1988}.}\label{tab:Final-PD-comparison}
\end{table}

Our value of the ${}^{1}\mathrm{H}^{-}$ photodetachment energy differs from the theoretical prediction reported by Drake \cite{DRAKE1988} by around $+5.9\times10^{-4}\,\mathrm{cm}^{-1}$. We have therefore carefully analyzed the result given in Table~3 of that work in order to identify the origin of this discrepancy, and reached the following conclusions. The largest deviation ($+7.7\times 10^{-4}\,\mathrm{cm}^{-1}$), originates from the approximate value of the Bethe logarithm adopted there, $\ln(k_{0}/\mathrm{Ry})=2.9718$, which is accurate to only two significant digits (see Table~\ref{tab:Bethe_log_results}). In addition, our relativistic correction differs by $-1.5\times10^{-4}\,\mathrm{cm}^{-1}$. We note that multiplying our nonrecoil relativistic contribution, $E_{\rm rel}=0.304\,397\,2\,\mathrm{cm}^{-1}$, by the reduced-mass factor $\mu/m$ reproduces exactly Drake’s value, suggesting a scaling error in the original evaluation. Finally, the remaining small deviation can be traced to several minor inaccuracies in the electron–nucleus and electron–electron terms of the QED corrections.

\begin{figure}[htbp]
\centering
\includegraphics[scale=0.45]{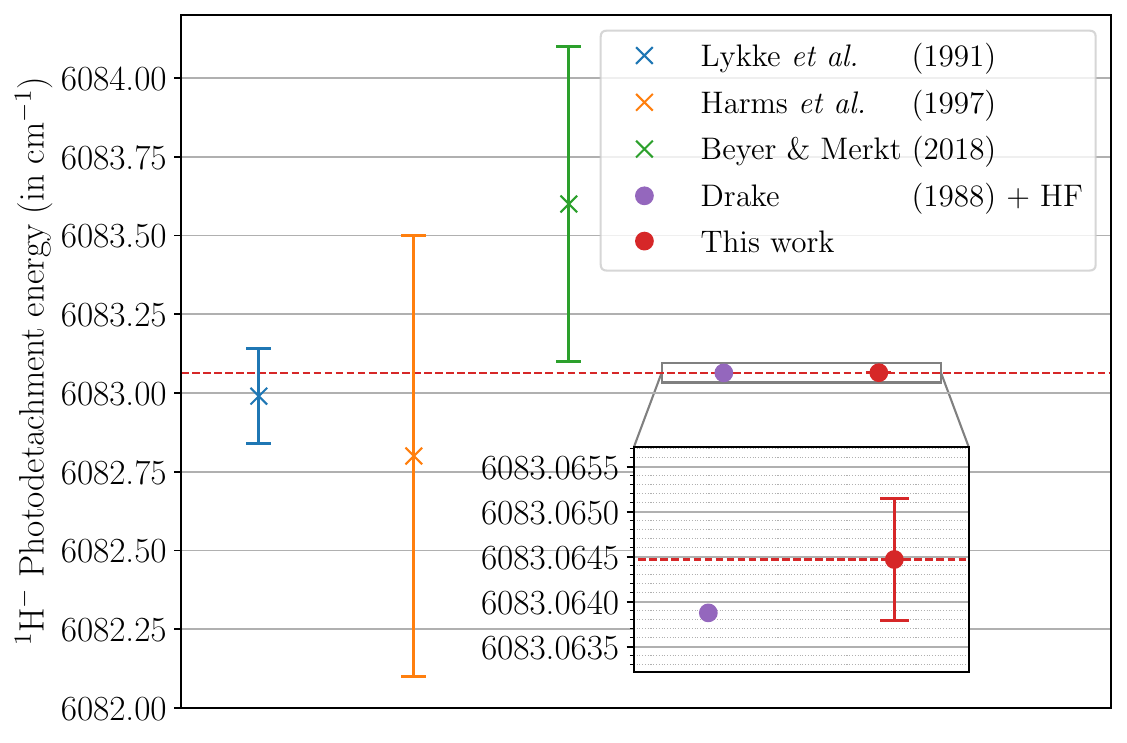}
\caption{\label{fig:Comparison-H} Comparison of experimental data (cross markers) and theoretical predictions (filled circles) of the ${}^1\mathrm{H}^{-}$ photodetachment energy (in $\mathrm{cm}^{-1}$). The horizontal red line extends our value for comparison. The HF correction was added to Drake's value (see Table \ref{tab:Final-PD-comparison}).}
\end{figure}

\begin{figure}[htbp]
\centering
\includegraphics[scale=0.45]{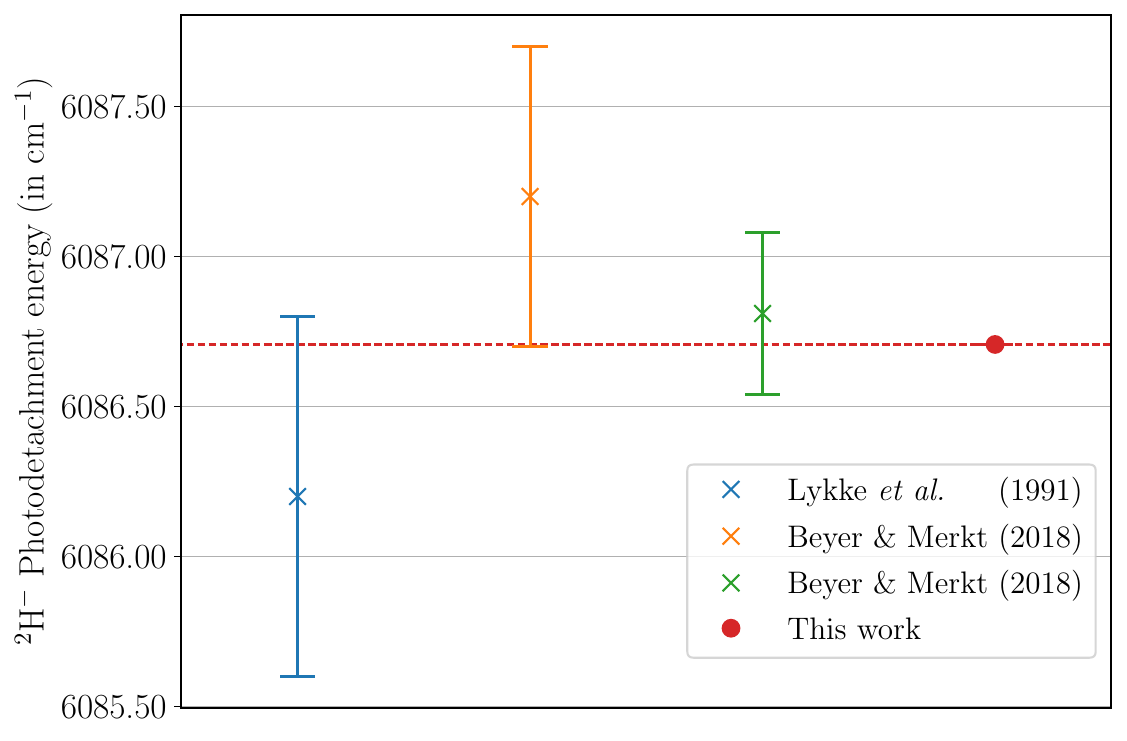}
\caption{\label{fig:Comparison-D} Comparison of experimental data (cross markers) and our theoretical prediction (filled circles) of the $^{2}\mathrm{H}^{-}$
photodetachment energy ($\mathrm{cm}^{-1}$). The horizontal red line extends our value for comparison.}
\end{figure}

The photodetachment energy obtained in the present work is in excellent agreement with the experimental determinations of Harms \textit{et al.} (1997) and Lykke \textit{et al.} (1991), thereby validating both measurements, as illustrated in Fig (\ref{fig:Comparison-H}). For the negative deuterium, $^{2}\mathrm{H}^{-}$, our result is compatible  with the three available experimental determinations while providing substantially higher precision, as shown in Fig. (\ref{fig:Comparison-D}). 

\section{Summary and outlook}

\label{sec:Summary-and-outlook}

We have performed a high-precision determination of the photodetachment
energy of the hydrogen anion $^{1}\mathrm{H}^{-}$, including explicit
electron correlation and recoil at the nonrelativistic level, together
with relativistic, QED, finite-nuclear-size, and hyperfine corrections.
The resulting value, $6083.06447(68)\,\mathrm{cm}^{-1}$, improves upon
previous theoretical and experimental determinations and establishes
a new reference for this benchmark system. Similar calculations for
$^{2}\mathrm{H}^{-}$ and $^{3}\mathrm{H}^{-}$ yield $6086.70679(68)$
and $6087.87924(68)$, respectively, in agreement with the available
experimental data for negative deuterium.

The achieved uncertainty of $3.1\times10^{-9}\,\mathrm{a.u.}$ ($0.084\,\mu\mathrm{eV}$)
surpasses the target accuracy of 1 $\mu\mathrm{eV}$ and is directly
relevant to ongoing experimental efforts. For the GBAR experiment,
a precise knowledge of the photodetachment threshold is essential,
to determine the laser energy required to produce ultracold $\bar{\mathrm{H}}^{+}$
ions. While photodetachment is not part of the current antihydrogen production schemes in the
ALPHA \cite{Alpha2023}, AEgIS \cite{AEgIS2018}, and ASACUSA \cite{Kuroda2014} experiments, these high-accuracy threshold values could serve as useful reference data for future schemes targeting the production of antihydrogen at low kinetic energies. Finally, we note that an improved experimental determination of the $\mathrm{H}^{-}$ photodetachment threshold at the 1 $\mu\mathrm{eV}$ precision level, which would allow testing the present calculations, is a realistic prospect~\cite{Ning2022,Carette2010}.

Further improvement beyond the present level of precision would require
the evaluation of the remaining contributions to the $H^{(4)}$ correction
as formulated by Pachucki \cite{Pachucki2006}. These include the
additional matrix elements listed in Table 1 of Ref. \cite{Pachucki2006},
as well as the $E_{A}^{\prime}$ contribution defined in Eq. (3.47)
of that work, which involves sums over the complete nonrelativistic
spectrum. Recoil effects at this order have been addressed in Ref.
\cite{PatkosYerokhinPachucki2017recoil}. The next contribution, $H^{(5)}$
of order $\alpha^{7}mc^{2}$, is currently known only for triplet
states \cite{PatkosYerokhinPachucki_triplet_alpha7_2021} and remains
unavailable for singlet states, as noted in Refs. \cite{YerokhinPatkosPachucki2021He,Patkos2025Habilitation}; given its high order in $\alpha$, its contribution lies well below the accuracy targeted in the present work.

\acknowledgements{We are grateful to V.I. Korobov for sharing his program for calculation of the Bethe logarithm in three-body systems, and for his helpful advice. We also thank Laurent Hilico, Vincent Barb\'e, and Mathis Panet for their careful reading of the manuscript. The authors acknowledge the support of the French Agence Nationale de la Recherche (ANR), under grant ANR-21-CE30-0047 (project Photoplus).}

\appendix

\section{Many-body Bethe logarithm}

\label{sec:Many-body-Bethe-logarithm}

In this appendix, we address the Bethe logarithm, the central and most intricate quantity that enters the leading-order QED correction in a general $N$-body quantum system with arbitrary masses and charges interacting via an instantaneous Coulomb potential. Our starting point is some state $|0\rangle$ of energy $E_{0}$, solving $H|n\rangle =E_{n}|n\rangle$, where the Hamiltonian reads
\begin{equation}
H=\sum_{i=0}^{N-1}\frac{\hat{\mathbf{P}}_{i}^{2}}{2m_{i}}+\sum_{i<j}^{N-1}\frac{q_{i}q_{j}}{r_{ij}}.\label{eq:many-body}
\end{equation}
The energy shift that the state $|0\rangle$ undergoes due to the self-energy
process is given by the following integral over the exchanged photon
energy $k$
\begin{equation}
\Delta E_{0}=\frac{2\alpha^{3}}{3\pi}\int_{0}^{\Lambda}kdk\langle 0|\mathbf{J}\cdot[E_{0}-H-k]^{-1}\mathbf{J}|0\rangle.\label{eq:de0}
\end{equation}
In this expression, the photon energy integral is regularized by a
cutoff parameter $\Lambda$, and $\mathbf{J}$ is the total charge-weighted
velocity operator, defined by 
\begin{equation}
\mathbf{J}=\sum_{i=0}^{N-1}q_{i}\hat{\mathbf{P}}_{i}/m_{i}.\label{eq:J}
\end{equation}
Note that Eq. (\ref{eq:de0}) yields two types of contributions: the
one-body self-energy processes (summed over all particles), corresponding
to paired terms with $\langle \mathbf{P}_{i}...\mathbf{P}_{i}\rangle$,
and the mutual single-photon exchange between two different particles
$i$ and $j$, corresponding to unpaired terms $\langle\mathbf{P}_{i}...\mathbf{P}_{j}\rangle$,
with $i\neq j$. We shall expand the resolvent operator present in
Eq. (\ref{eq:de0}), in the complete set of solutions of Eq. (\ref{eq:many-body}),
as
\begin{equation}
[E_{0}-H-k]^{-1}=\sum_{n}(E_{0}-E_{n}-k)^{-1}|n\rangle\langle n|.\label{eq:resolvent}
\end{equation}
Moreover, to isolate the divergences of our last integral,
one can write the energy factor as
\begin{equation}
\frac{1}{E_{0}-E_{n}-k}=-\frac{1}{k}+\frac{1}{k}\frac{E_{n}-E_{0}}{E_{n}-E_{0}+k}.
\end{equation}
The first term in this expression yields a linear divergence, which
is eradicated by the mass renormalization procedure \cite{Bethe1947},
and the second term yields a logarithmic divergence, together with
a finite physical contribution. As it was first shown by Bethe, the
mass renormalization procedure is performed through subtracting the
free-electron self-energy contribution, $\Delta E_{0}^{\mathrm{SE}}=\Delta E_{0}-\Delta E_{0}^{\mathrm{Free}}$,
yielding
\begin{align}
\Delta E_{0}^{\mathrm{SE}} & =\frac{2\alpha^{3}}{3\pi}\sum_{n}\int_{0}^{\Lambda}dk\frac{E_{n}-E_{0}}{E_{n}-E_{0}+k}|\langle 0|\mathbf{J}|n\rangle |^{2}.
\end{align}
This integral is to be evaluated in the sense of Cauchy principal
value, giving
\begin{equation}
\mathrm{PV}\int_{0}^{\Lambda}\frac{dk}{E_{n}-E_{0}+k}=\ln\frac{\Lambda}{|E_{n}-E_{0}|}+{\cal O}(\Lambda^{-1}).
\end{equation}
These results yield the cutoff-dependent self-energy shift
\begin{align}
\Delta E_{0}^{\mathrm{SE}} & =\frac{2\alpha^{3}}{3\pi}\ln(\Lambda/k_{0}){\cal D}\label{eq:SE_log_lambda}
\end{align}
In this expression appears the Bethe logarithm $\ln(k_{0}/\mathrm{Ry})={\cal N}/{\cal D}$,
where the numerator ${\cal N}$ and denominator ${\cal D}$ parts
are given by
\begin{align}
{\cal N} & =\sum_{n}(E_{n}-E_{0})\ln(|E_{n}-E_{0}|/\mathrm{Ry})|\langle 0|\mathbf{J}|n\rangle | ^{2}\label{eq:N_sum}\\
{\cal D} & =\sum_{n}(E_{n}-E_{0})|\langle 0 |\mathbf{J} |n\rangle |^{2}
,\label{eq:D_sum}
\end{align}
respectively. While the denominator can be easily evaluated, since it reduces to a sum over expectation values of Dirac delta functions \cite{Korobov2004PhysRevA.70.012505}
\begin{equation}
\begin{aligned}
{\cal D} & =\langle 0 |\mathbf{J}\cdot(H-E_{0})\mathbf{J} |0\rangle=\frac{1}{2}\langle 0 |[\mathbf{J},[H,\mathbf{J}]]|0\rangle\\
 & =-2\pi\sum_{j>i=0}^{N-1}q_{i}q_{j}\left(\frac{q_{i}}{m_{i}}-\frac{q_{j}}{m_{j}}\right)^{2}\langle 0|\delta(\mathbf{r}_{ij}) |0\rangle,
\end{aligned}
\label{eq:denominator_expansion}
\end{equation}
the evaluation of the numerator ${\cal N}$ is more challenging. As it was
noted by Schwartz \cite{Schwartz1961}, as well as Drake and Goldman
\cite{DrakeGoldman2000}, the direct evaluation of ${\cal N}$ in
Eq. (\ref{eq:N_sum}) is problematic due to the poor representation
of high-energy solutions in the finite basis approximation. The solution is to rewrite it as \cite[Eq. (7)]{Schwartz1961}\cite[Eqs. (3,4)]{DrakeGoldman2000}
\begin{equation}
\begin{aligned}\!{\cal N} \!=\!\!\lim_{\Lambda\rightarrow\infty}\!\!\big[\!-\!\Lambda\langle 0|\mathbf{J}^{2}|0\rangle\!+\!\ln(\Lambda/\mathrm{Ry}){\cal D}\!+\!\int_{0}^{\Lambda}\!\!\!\!\!kdkJ(k)\big].\label{eq:NSchwartz}
\end{aligned}
\end{equation}
In this expression appears the function $J(k)$, given by
\begin{align}
J(k) & =\langle 0|\mathbf{J}\cdot [H-E_{0}+k]^{-1}\mathbf{J}|0\rangle,\label{eq:Jk}
\end{align}
and the first two terms cancel the linear and logarithmic singularity
arising from the last integral, in the limit $\Lambda\rightarrow\infty$, yielding the finite physical result.

We follow the approach of Korobov \cite{Korobov2012}, which builds
on the foundational work of Schwartz \cite{Schwartz1961}. While Schwartz
obtained a benchmark evaluation of the helium Lamb shift in the infinite
nuclear mass limit, Korobov extended the formalism to few-body Coulomb
systems with arbitrary masses and charges, thus including recoil effects
non-perturbatively. Crucially, the high accuracy of Korobov\textquoteright s
results stems from the use of an optimized exponential basis set for
the numerical evaluation of the Bethe logarithm, leading to the most
precise helium values available to date \cite{Korobov2019,Yang2025}.
Following the cited works, we write the last integral as
\begin{equation}
\int_{0}^{\Lambda}kdkJ(k)=\int_{0}^{K}kdkJ(k)+\int_{K}^{\Lambda}kdkJ(k),\label{eq:J_cuts}
\end{equation}
split into two regions of the photon energy. The integral along the
first segment, $k\in[0,K]$ (low-energy), is evaluated analytically,
\begin{equation}
\begin{aligned} & \mathrm{PV}\int_{0}^{K}kdkJ(k) =\\
\! &\sum_{n}|\langle 0|\mathbf{J}|n\rangle |^{2} \bigg[K\! -\!(E_{0}\!-\!E_{n})\ln \bigg|\frac{E_{0}\!-\!E_{n}}{E_{0}\!-\!E_{n}\!-\!K}\bigg|\bigg],
\end{aligned}\label{eq:low-energy}
\end{equation}
and the sum shall be carried along all numerical eigen-solutions of
the many-body problem of Eq. (\ref{eq:many-body}). With an adequate
basis set this quantity can be very well converged. Moreover, the
function $J(k)$ follows a maximization variational principle. At
each point $k$, we have $J_{\mathrm{trial}}(k)\leq J_{\mathrm{exact}}(k)$
\cite{Korobov2012}. This means that one could optimize the basis
set (exponents) of the intermediate states $|n\rangle$ to
maximize the integral of Eq. (\ref{eq:low-energy}), for some values
of $K$, typically at $\sim10^{i} \, \mathrm{ a.u.}$ with $i=3,4,5$. For
the high-energy segment, $k\in[K,\Lambda]$, where the sum over states
is no longer efficient, for a finite basis set, analytical asymptotic
expressions for $J(k)$ can be employed,
\begin{equation}
\begin{aligned}J(k)= & \frac{\langle 0 |\mathbf{J}^{2} |0\rangle}{k}-\frac{{\cal D}}{k^{2}}+J_{3}(k)+J_{\mathrm{rem}}(k).\end{aligned}
\label{eq:Jk_expansion}
\end{equation}
While obtaining the first two terms is straightforward, the
calculation of the last two terms is more involved. The $J_{3}$ function
was derived by Schwartz, Korobov, in addition to Forrey and Hill
\cite{Schwartz1961,KorobovKorobov1999,Korobov2012,ForreyHill1993}
\begin{equation}
J_{3}(k)=\frac{A_{3}\sqrt{k/m}+B_{3}\ln(k/m)+C_{3}}{k^{3}},\label{eq:J3}
\end{equation}
and the three corresponding coefficients for the many-body problem
were found to be 
\begin{equation}
\begin{aligned}A_{3} & =4\pi\sum_{i>j}q_{i}^{2}q_{j}^{2}\left(\frac{q_{i}}{m_{i}}\!-\!\frac{q_{j}}{m_{j}}\right)^{2}\sqrt{2m_{ij}m}\langle \delta(\mathbf{r}_{ij})\rangle,\\
B_{3} & =4\pi\sum_{i>j}q_{i}^{3}q_{j}^{3}\left(\frac{q_{i}}{m_{i}}\!-\!\frac{q_{j}}{m_{j}}\right)^{2}m_{ij}\langle\delta(\mathbf{r}_{ij})\rangle,\\
C_{3} &= \hspace{-3mm} \sum_{\substack{i>j,k>l\\
(i,j)\neq(k,l)}}\!\! q_{i}q_{j}q_{k}q_{l}\left(\frac{q_{i}}{m_{i}}\!-\!\frac{q_{j}}{m_{j}}\right)\!\!\!\left(\frac{q_{k}}{m_{k}}\!-\!\frac{q_{l}}{m_{l}}\right)\!\!\langle\frac{\mathbf{r}_{ij}\!\cdot\!\mathbf{r}_{kl}}{r_{ij}^{3}r_{kl}^{3}}\rangle\\
 & +4\pi\sum_{i>j}q_{i}^{2}q_{j}^{2}\left(\frac{q_{i}}{m_{i}}\!-\!\frac{q_{j}}{m_{j}}\right)^{2}\bigg\{\mathcal{R}_{ij}\!+\!q_{i}q_{j}m_{ij}[-\ln2\\
 & \qquad-1+\ln(m_{ij}/m)]\langle\delta(\mathbf{r}_{ij})\rangle\bigg\}. \label{eq:A3B3C3}
\end{aligned}
\end{equation}
Here, $m_{ij}\equiv m_{i}m_{j}/(m_{i}+m_{j})$, and the matrix element $\mathcal{R}_{ij}$
is defined in \cite[Eq. (20)]{Korobov2012}. The remainder function
$J_{\mathrm{rem}}(k)$, shall be later discussed. If one now plugs Eq.
(\ref{eq:J3}) into Eq. (\ref{eq:NSchwartz}), one finds
\begin{equation}
\begin{aligned}{\cal N} & =\int_{0}^{K}kdkJ(k)-K\langle 0|\mathbf{J}^{2}|0\rangle-\ln(\mathrm{Ry}/K){\cal D}\\
 & +\int_{K}^{\infty}kdk\left\{ J_{3}(k)+J_{\mathrm{rem}}(k)\right\} ,
\end{aligned}
\end{equation}
the remaining function, $J_{\mathrm{rem}}(k)$, shall be written as
a sum
\begin{equation}
J_{\mathrm{rem}}(k)=\sum_{n=4}^{n_{\mathrm{max}}}\frac{A_{n}\sqrt{k/m}+B_{n}\ln(k/m)+C_{n}}{k^{n}}, \label{eq:Jrem}
\end{equation}
having a structure similar to that $J_{3}(k)$, as well as the structure of the
corresponding analytical $J_{\mathrm{rem}}(k)$ of the hydrogenic ground
\cite[Eqs. (A1, A2)]{Korobov2012}\cite[sec. 3.3.3]{BukowskiJeziorskiMoszynskiKolosqua.560420205}
and excited states \cite[Table I \& Eqs. (52-56)]{Maquet1977PhysRevA.15.1088}.
The coefficients in this expansion shall be obtained by least-square-fitting
of this function to the values of $J(k)-\langle 0 |\mathbf{J}^{2} |0\rangle/k+{\cal D}/k^{2}-J_{3}(k)$,
computed in some range $k\in[k_{\mathrm{min}},k_{\mathrm{max}}]$, where
$k_{\mathrm{max}}\leq K$, a range on which $J(k)=\sum_{n}[E_{n}-E_{0}+k]^{-1}|\langle 0|\mathbf{J}|n\rangle |^{2}$,
entering the low-energy integral, is expected to be very accurately
represented (optimized through maximization). The number of coefficients
$3(n_{\mathrm{max}}-3)$ is varied, together with the energy parameters
$K$, $k_{\mathrm{min}}$, and $k_{\mathrm{max}}$, until stabilization,
and an overall fitting error bar is provided. Once these coefficients
are obtained, the high-energy integral is evaluated analytically,
and one is left with the final expression of the Bethe logarithm
\begin{equation}
\begin{aligned} & \ln(k_{0}/\mathrm{Ry})=\frac{1}{{\cal D}}\bigg[\int_{0}^{K}kdkJ(k)-K\langle 0|\mathbf{J}^{2}|0\rangle\\
 & -\ln(\mathrm{Ry}/K){\cal D}+\sum_{n=3}^{n_{\mathrm{max}}}K^{2-n}\bigg\{ A_{n}\frac{2\sqrt{K/m}}{2n-5}\\
 & +B_{n}\frac{(n-2)\ln(K/m)+1}{(n-2)^{2}}+\frac{C_{n}}{n-2}\bigg\}\bigg].
\end{aligned}\label{eq:Bethe}
\end{equation}

It is worth noting that the coefficient $A_{4}$ has been derived
by Forrey and Hill \cite[Eq. (3.5)]{ForreyHill1993}, for the case
of $N$ electrons and $K$ infinite-mass nuclei. Although we have
not incorporated their closed-form result into our expansion, using
it directly rather than determining $A_{4}$ through fitting would
improve the accuracy of Bethe-logarithm calculations. Their approach
also provides a viable route to obtain higher-order coefficients ($B_{4},C_{4},...$),
which, to our knowledge, have not been pursued in the literature.
Given the limited visibility of this work, further development along
their lines would be of clear value to the field. We finally note
that the quality of $\ln(k_{0}/\mathrm{Ry})$ can always be enhanced
by increasing $K$, which suppresses the high-energy contributions
of the fitted terms ($A_{i}$ $B_{i}$ and $C_{i}$, with $i\geq4$,
and their uncertainties) through the factor $K^{2-n}$. The drawback
is the need for a substantially larger basis set to converge the low-energy
integral to the required precision, as noted in Ref. \cite[p. 2]{KorobovZhong2012}. 

We have so far formulated our problem for the general many-body problem
where the Hamiltonian is given by Eq. (\ref{eq:many-body}), and the
corresponding Bethe logarithm is computed through Eq. (\ref{eq:Bethe}).
On the other hand, in the center-of-mass frame, the Bethe logarithm
calculation, just as the nonrelativistic Schr{\"o}dinger calculation,
is further simplified by eliminating the center-of-mass degree of
freedom. With the change of variables from $\{\mathbf{R}_{0},\ldots,\mathbf{R}_{N-1}\}$,
to $\{\mathbf{G},\mathbf{r}_{1},\ldots,\mathbf{r}_{N-1}\}$,
where the center of mass $\mathbf{G}$ and the relative coordinate
$\mathbf{r}_{i}$ are given by
\begin{align}
\mathbf{G} & =\frac{1}{M_{\mathrm{tot}}}\sum_{i=0}^{N-1}m_{i}\mathbf{R}_{i},\quad\text{with}\quad M_{\mathrm{tot}}=\sum_{i=0}^{N-1}m_{i},\\
\mathbf{r}_{i} & =\mathbf{R}_{i}-\mathbf{R}_{0},\quad\text{with}\quad i=1,\ldots,N-1,
\end{align}
one can write the total $N$-body Hamiltonian of Eq. (\ref{eq:many-body})
as $H=\tilde{H}+H_{G}$, with
\begin{align}
\tilde{H} & =\sum_{i=1}^{N-1}\frac{\hat{\mathbf{p}}_{i}^{2}}{2\mu_{i}}+\frac{1}{m_{0}}\sum_{i>j\geq1}^{N-1}\hat{\mathbf{p}}_{i}\cdot\hat{\mathbf{p}}_{j}\nonumber \\
 & +\sum_{i=1}^{N-1}\frac{q_{i}q_{0}}{r_{i}}+\sum_{i>j\geq1}^{N-1}\frac{q_{i}q_{j}}{r_{ij}}\\
H_{G} & =\frac{\hat{\mathbf{p}}_{G}^{2}}{2M_{\mathrm{tot}}},
\end{align}
where the motion of the center-of-mass is decoupled from the internal
motion of relative particles. At the level of states, one can define
$|n\rangle=|\tilde{n}\rangle \otimes |n_{G}\rangle$, where $|\tilde{n}\rangle$
and $|n_{G}\rangle$, are decoupled normalized states, solving $\tilde{H}|\tilde{n}\rangle=\tilde{E}_{n}|\tilde{n}\rangle$,
and $H_{G}|n_{G}\rangle=E_{n,G}|n_{G}\rangle$, respectively.
In the center-of-mass frame, we have $E_{n}=\tilde{E}_{n}$, since
$E_{n,G}$ vanishes. Using this construction, one can show that the
evaluation of the Bethe logarithm through Eq. (\ref{eq:Bethe}) is
equivalent to evaluating it in 
\begin{equation}
\begin{aligned} & \ln(k_{0}/\mathrm{Ry})=\frac{1}{\tilde{{\cal D}}}\bigg[\int_{0}^{K}kdk\tilde{J}(k)-K\langle \tilde{0}|\tilde{\mathbf{J}}^{2}|\tilde{0}\rangle\\
 & -\ln(\mathrm{Ry}/K)\tilde{{\cal D}}+\sum_{n=3}^{n_{\mathrm{max}}}K^{2-n}\bigg\{ A_{n}\frac{2\sqrt{K/m}}{2n-5}\\
 & +B_{n}\frac{(n-2)\ln(K/m)+1}{(n-2)^{2}}+\frac{C_{n}}{n-2}\bigg\}\bigg],
\end{aligned}\label{eq:BetheTilde}
\end{equation}
where $\tilde{J}(k)\equiv\langle\tilde{0}|\tilde{\mathbf{J}}\cdot[\tilde{H}-\tilde{E_{0}}+k]^{-1}\tilde{\mathbf{J}}|\tilde{0}\rangle$,
and where everything is formulated in terms of internal variables.
This comes from the fact that one can easily show that $\tilde{{\cal D}}=\langle\tilde{0}|\tilde{\mathbf{J}}[\tilde{H}-\tilde{E}_{0}]^{-1}\tilde{\mathbf{J}}|\tilde{0}\rangle={\cal D}$,
and $\langle\tilde{0}|\tilde{\mathbf{J}}^{2}|\tilde{0}\rangle=\langle 0 |\mathbf{J}^{2}|0\rangle$.
In the new system of coordinates our $\mathbf{J}$ operator reads
\begin{equation}
\mathbf{J}=\sum_{i=1}^{N-1}\left[\frac{q_{i}}{m_{i}}-\frac{q_{0}}{m_{0}}\right]\hat{\mathbf{p}}_{i}+\frac{Q_{\mathrm{tot}}}{M_{\mathrm{tot}}}\hat{\mathbf{p}}_{G},
\end{equation}
where the first term is $\tilde{\mathbf{J}}$, and $Q_{\mathrm{tot}}$
is the total charge of the system. The only difference between the
two problems is the fact that $\tilde{J}(0)\neq J(0)$, while the
two functions coincide for all $k>0$. To show that the first statement
is true, we introduce the operator $\mathbf{O}=\sum_{k=0}^{N-1}iq_{k}\mathbf{R}_{k},$
which satisfies $[H,\mathbf{O}]=\mathbf{J}$, where $H$ and
$\mathbf{J}$ are given in Eqs (\ref{eq:many-body} and \ref{eq:J}).
One can then easily show that
\begin{equation}
\begin{aligned}J(0) & =\langle 0|\mathbf{J}[H-E_{0}]^{-1}\mathbf{J}|0\rangle\\
 & =-\langle 0 |[\mathbf{O},[H,\mathbf{O}]] |0 \rangle /2=\frac{3}{2}\sum_{i=0}^{N-1}\frac{q_{i}^{2}}{m_{i}}.
\end{aligned}\label{eq:J0}
\end{equation}

Similarly, for the tilde problem of relative coordinates, one can
introduce the corresponding operator $\tilde{\mathbf{O}}=\sum_{i=1}^{N-1}i(q_{i}-Q_{\mathrm{tot}}m_{i}/M_{\mathrm{tot}})\mathbf{r}_{i},$
which in turn satisfies the commutation relation $[\tilde{H},\tilde{\mathbf{O}}]=\tilde{\mathbf{J}}$,
and allows one to find
\begin{equation}
\begin{aligned} & \tilde{J}(0) =\langle\tilde{0}|\tilde{\mathbf{J}}\cdot[\tilde{H}-\tilde{E}_{0}]^{-1}\tilde{\mathbf{J}}|\tilde{0}\rangle\\
 & =-\langle\tilde{0}|[\tilde{\mathbf{O}},[\tilde{H},\tilde{\mathbf{O}}]]|\tilde{0}\rangle/2=\frac{3}{2}\big[\sum_{i=0}^{N-1}\frac{q_{i}^{2}}{m_{i}}-\frac{Q_{\mathrm{tot}}^{2}}{M_{\mathrm{tot}}}\big],
\end{aligned}\label{eq:Jtilde0}
\end{equation}
a result that is consistent with that of \textit{et al.} in
\cite[Eq. (20)]{ZhouZhuYan2006}. To validate these results,
and show that $J(k)$ coincides with $\tilde{J}(k)$ for all $k>0$,
we formulate the first function in terms of the second one and find
\begin{equation}
\begin{aligned}
 & J(k)= \\
 & \big(\frac{Q_{\mathrm{tot}}}{M_{\mathrm{tot}}}\big)^{2}\langle 0_{G}|\hat{\mathbf{p}}_{G}\!\cdot\![E_{0,G}\!-\!\frac{\hat{\mathbf{p}}_{G}^{2}}{2M_{\mathrm{tot}}}\!-\!k]^{-1}\hat{\mathbf{p}}_{G}|0_{G}\rangle\\
 & \!+2\frac{Q_{\mathrm{tot}}}{M_{\mathrm{tot}}}\langle 0_{G}|\hat{\mathbf{p}}_{G}\!\cdot\![E_{0,G}\!-\!\frac{\hat{\mathbf{p}}_{G}^{2}}{2M_{\mathrm{tot}}}\!-\!k]^{-1}|0_{G}\rangle \langle \tilde{0} |\tilde{\mathbf{J}} |\tilde{0}\rangle\\
 & +\langle\tilde{0} |\tilde{\mathbf{J}}\!\cdot\! [\tilde{E}_{0}\!-\!\tilde{H}\!-\!k]^{-1}\tilde{\mathbf{J}} |\tilde{0}\rangle.
\end{aligned}
\end{equation}
For a general $k>0$, and in the center-of-mass frame one can easily
show that the first term vanishes, and the first expectation value
of the second line vanishes, in addition to the vanishing of $\langle\tilde{0} |\tilde{\mathbf{J}} |\tilde{0}\rangle=\langle\tilde{0}|[H,\tilde{\mathbf{O}}]|\tilde{0}\rangle=(E_{0}-E_{0})\langle\tilde{0}|\tilde{\mathbf{O}}|\tilde{0}\rangle=0$.
The last term is nothing but $\tilde{J}(k)$. For the special case
of $k=0$, the first two terms become singular, due to the action
of $[\hat{\mathbf{p}}_{G}^{2}]^{-1}$ on a state with vanishing
momentum. We circumvent this inconvenience by introducing a small
harmonic regularizer through the replacement
\begin{equation}
\frac{\hat{\mathbf{p}}_{G}^{2}}{2M_{\mathrm{tot}}}\rightarrow\frac{\hat{\mathbf{p}}_{G}^{2}}{2M_{\mathrm{tot}}}+\frac{1}{2}M_{\mathrm{tot}}\omega\mathbf{G}^{2},
\end{equation}
where $\omega$ is some small frequency, allowing the parametrization
of the singularity of $[\hat{\mathbf{p}}_{G}^{2}]^{-1}$, as well
as the discretization of its spectrum. This regularization, with further
straightforward harmonic oscillator algebra, yields the following
connection between the two functions under consideration
\begin{equation}
J(k)=-\frac{3}{2}\frac{Q_{\mathrm{tot}}^{2}}{M_{\mathrm{tot}}}\lim_{\omega\rightarrow0}\left(\frac{\omega}{\omega-k}\right)+\tilde{J}(k),
\end{equation}
showing that for non-vanishing $k$, the first term vanishes, and
we obtain $J(k)=\tilde{J}(k)$, validating our previous observation.
For vanishing $k$, the first term yields $-(3/2)Q_{\mathrm{tot}}^{2}/M_{\mathrm{tot}}$,
a result that is consistent with our previous results in Eqs. (\ref{eq:J0})
and (\ref{eq:Jtilde0}). Note that for the case of neutral systems,
or a system where a single mass is taken to be infinite, this extra
term vanishes for all $k$. We note that when evaluating the low-energy
integral, this extra contribution vanishes. In our actual numerical
evaluation, we shall work with the tilde problem in Eq. (\ref{eq:BetheTilde}),
where the value of $\tilde{J}(0)$ will give a rough estimation of
our numerical precision for $\tilde{J}(k)$.

\bibliography{bibliography}

\end{document}